\documentclass[aps,showpacs,amsmath,amssymb,pra,twocolumn,superscriptaddress]{revtex4}
\usepackage{graphicx}
\usepackage{bm}
\usepackage{color}
\definecolor{mygrey}{rgb}{.9,.9,.9}
\begin{document}
\title{Ultracold Rydberg Atoms in a Ioffe-Pritchard Trap}
\date{\today}

\author{Bernd Hezel}
\email[]{hezel@physi.uni-heidelberg.de}
\affiliation{Physikalisches Institut, Universit\"at Heidelberg, Philosophenweg 12, 69120 Heidelberg, Germany}

\author{Igor Lesanovsky}
\email[]{Igor.Lesanovsky@uibk.ac.at}
\affiliation{Institut f\"ur Theoretische Physik, Universit\"at Innsbruck, Technikerstr. 25, 6020 Innsbruck, Austria}

\author{Peter Schmelcher}
\email[]{Peter.Schmelcher@pci.uni-heidelberg.de}
\affiliation{Physikalisches Institut, Universit\"at Heidelberg, Philosophenweg 12, 69120 Heidelberg, Germany}
\affiliation{Theoretische Chemie, Physikalisch-Chemisches Institut, Universit\"at Heidelberg, INF 229, 69120 Heidelberg, Germany}%

\begin{abstract}
We discuss the properties of ultracold Rydberg atoms in a Ioffe-Pritchard magnetic field configuration. 
The derived two-body Hamiltonian unveils how the large size of Rydberg atoms affects their coupling to the inhomogeneous magnetic field. 
The properties of the compound electronic and center of mass quantum states are thoroughly analyzed. 
We find very tight confinement of the center of mass motion in two dimensions to be achievable  
while barely changing the electronic structure compared to the field free case. 
This paves the way for generating a one-dimensional ultracold quantum Rydberg gas. 
\end{abstract}

\pacs{32.60.+i,33.55.Be,32.10.Dk,33.80.Ps}
\maketitle

 \section{\label{s:intro}Introduction}

Powerful experimental cooling techniques have been developed in the past decades that
allow us to probe the micro and nanokelvin regime while controlling 
the internal and external degrees of freedom of atomic systems.
As a result dilute ultracold gases that qualify perfectly for 
the study of quantum phenomena on a macroscopic scale \cite{pethick01,dalfovo99,pitaevskii03}
can nowadays be prepared almost routinely.
Although being dilute, interactions play an important role and
rich collective phenomena, reminiscent of e.g. those in traditional condensed matter physics, appear.

The attractiveness of Rydberg atoms arises from their extraordinary properties \cite{gallagher}.
The large displacement of the valence electron and the atomic core 
is responsible for the exaggerated response to external fields and, therewith, for their enormous polarizability.
Rydberg atoms possess large dipole moments and, despite being electronically highly excited, 
they can possess lifetimes of the order of milliseconds or even more.

Due to their susceptibility with respect to external fields and/or their long range
interaction, ensembles of Rydberg atoms represent intriguing many-body systems with
rich excitations and decay channels. Starting from laser cooled ground state atoms, a laser typically
excites a subensemble of the atoms to the desired Rydberg states.
Since the ultraslow motion of the atoms can be ignored on short timescales, 
Rydberg-Rydberg interactions dominate the system and we encounter a so-called frozen
Rydberg gas \cite{mourachko}. The strength of the interaction can be varied by tuning external fields
and/or by selecting specific atomic states. An exciting objective are the many-body effects
to be unraveled in ultracold Rydberg gases (see Refs. \cite{pohl1,pohl2} and references therein).
At a certain stage of the evolution ionization might take over leading to a cold Rydberg plasma. 

Beyond the above there is a number of topical and promising research activities involving
cold Rydberg states. One example are long range molecular Rydberg states \cite{greene}
with unusual properties if exposed to magnetic fields \cite{igor:jpb39}. 
Another one is due to the strong dipole-dipole interaction of Rydberg atoms which strongly inhibits excitation
of their neighbors \cite{singer,tong}.  The resulting local excitation blockade is state dependent 
and can turn Rydberg atoms into possible candidates for 
quantum information processing schemes~\cite{ryabtsev,lukin}. 
 
A precondition for enabling the processing of Rydberg atoms 
is the availability of tools to control their quantum behavior and properties.
An essential ingredient in this respect is the trapping of electronically highly excited atoms.
The present work provides a major contribution on this score. Let us briefly address
previous works on Rydberg atoms exposed to inhomogeneous static field configurations.

First evidence for trapped Rydberg gases has been experimentally found by Choi~et~al.~\cite{choi,choi:epjd}. 
The authors use strong bias fields to trap ``guiding center'' drift atoms for up to $200$~ms. 
Quantum mechanical studies of highly excited atoms in magnetic quadrupole fields demonstrated
the existence of e.g. intriguing spin polarization patterns and magnetic field-induced
electric dipole moments \cite{igor:epl,igor:jpb}. These investigations were based on the 
assumption of an infinitely heavy nucleus. A description of the coupled center of mass (c.m.) and
electronic dynamics has been presented in Refs.\cite{igor:prl,igor:pra72}: 
Trapping has been achieved for quantum states with sufficiently large total, i.e.\ electronic
and c.m., angular momentum.  Pictorially speaking this addresses atoms that circle
around the point of zero field at a sufficiently large distance.
Recently it has been demonstrated that trapping in a Ioffe-Pritchard configuration
is possible without imposing the condition of large c.m. angular momenta \cite{hezel:prl}. 
The present investigation works out this setup in detail and provides comprehensive results
for Rydberg atoms exposed to the Ioffe-Pritchard field configuration.

In detail we proceed as follows.
Sect.~\ref{s:h} contains a derivation of our working Hamiltonian for a highly excited 
atom in the inhomogeneous field including the coupling of the electronic and c.m.\ motion of the atom. 
In Sect.~\ref{s:aa} we introduce an adiabatic approximation in order to solve the corresponding
stationary Schr\"odinger equation. In Sect.~\ref{s:ees} we analyze the obtained spectra and
point out the capacity of the Ioffe bias field 
to regulate the distance between the surfaces, and with that the quality of the adiabatic approach. 
Intersections through the surfaces show their deformation when the field gradient is increased. 
Subsequently we characterize the electronic wave functions by discussing relevant expectation values.
Sect.~\ref{s:cm} is dedicated to the c.m. dynamics in the uppermost adiabatic energy surface.
We arrive at a confined quantized c.m. motion without the need to impose any restriction
on its properties. Examining the fully quantized states we observe that 
the extension of the electronic cloud can exceed the extension of the c.m.\ wave function. 

\section{\label{s:h}Hamiltonian}

\subsection{\label{s:h:tbh}Two-body Approach}

The large distance of the highly excited
valence electron (particle 1) from the remaining closed-shell ionic core of an alkali Rydberg atom (particle 2)
renders it possible to model the mutual interaction by an effective potential 
which is assumed to depend only on the distance of the two particles.
For alkali atoms, in particular, 
whose core possess zero total angular momentum and zero total spin, 
the only essential difference to the Coulombic case is due to the finite size of the core.
In any case, the effective potential $V(r)$ only noticeably differs 
from the pure Coulomb potential at small distances~$r$.
States of high \emph{electronic} angular momenta $l$, 
on which we focus in the present investigation, 
almost exclusively probe the Coulombic tail of this potential.

The coupling of the charged particles to the external magnetic field 
is introduced via the minimal coupling, 
$\bm{p}\rightarrow\bm{p}-q \bm{A}$, where $q$ is the charge of the particle and $\bm{A}$ is a vector potential 
belonging to the magnetic field $\bm{B}$.  Including the coupling of the magnetic moments to the external field 
($\bm{\mu}_1$ and $\bm{\mu}_2$ originate from the electronic and nuclear spin, respectively), 
our initial Hamiltonian reads 
(we use atomic units except when stated otherwise) 
\begin{eqnarray} \label{eq:Hinit}
    H_{init}&=&\frac{1}{2M_1}\left(\bm{p}_1-q_1\bm{A}(\bm{r}_1)\right)^2
    +\frac{1}{2M_2}(\bm{p}_2-q_2\bm{A}(\bm{r}_2))^2 \nonumber\\
    &&+V(\left|\bm{r}_1-\bm{r}_2\right|)
    -\bm{\mu}_1\cdot\bm{B}(\bm{r}_1)-\bm{\mu}_2\cdot\bm{B}(\bm{r}_2) \; .
\end{eqnarray}
We do not take into account spin-orbit-coupling and relativistic mass changes. 
The difference in energy shift for adjacent, large angular momentum states ($l$, $l\pm 1$)  
due to these relativistic corrections 
is $\Delta W_{FS}=\alpha^2/2n^5$ \cite{bethe}, 
where $\alpha$ is the fine structure constant, and therefore negligible for Rydberg states.
At $n=30$ one receives $\Delta W_FS=1.1\times 10^{-12}$ atomic units.
To give an idea of the scope of this approximation we anticipate a result from Sec.~\ref{s:ees}: 
The energy gap between two adjacent high-l electronic states is approximately $E_{dist}=B/2$ a.u. 
Demanding $\Delta W_{FS} / E_{dist} \ll 1$ results is constraining the Ioffe field strength $B$ 
to be much larger than $5$ mG.

Before we focus on the Ioffe-Pritchard configuration 
let us first examine a general field  $\bm B$ composed of 
a constant term $\bm B_c$, a linear term $\bm B_l$ and higher order terms, $\bm{B}=\sum \bm{B}_i$.
The vector potential shall satisfy the Coulomb gauge. 
The squared terms can then be simplified taking advantage of the vanishing commutator 
$[\bm{A}(\bm{r}_1),\bm{p}_1]$ to obtain 
$(\bm{p}_1-q \bm{A}(\bm{r}_1))^2= \bm{p}_1^2-2 q \bm{A}(\bm{r}_1)\cdot\bm{p}_1+ q^2 \bm{A}(\bm{r}_1)^2$.
In the so-called symmetric gauge 
the vector potential of a constant magnetic field is given by
$\bm{A}_c(\bm{r}_1)=1/2\: \bm{B}_c\times\bm{r}_1$. 
The analogon for a linear field is 
$\bm{A}_l(\bm{r}_1)=1/3\: \bm{B}_l(\bm{r}_1)\times\bm{r}_1$.
It can be proven that the vector potential of an arbitrary magnetic field can be expanded
in a corresponding form \cite{igor:diss}
permitting a representation of the vector potential as a cross product 
$\bm{A}(\bm{r}_1) = \sum_i \bm{A}_i(\bm{r}_1) 
=\bm{\tilde{B}}(\bm{r}_1)\times\bm{r}_1$,
where $\bm{\tilde{B}}(\bm{r}_1)=\sum g_i \bm{B}_i(\bm{r}_1)$ 
and $i\in \{c,l,\dots\}$ denotes the order of the corresponding terms of $\bm A$ and $\bm B$ 
with respect to spacial coordinates. 
$g_i$ are the coefficients $\frac{1}{2}$, $\frac{1}{3}$ etc.
The particular form of this potential and the vanishing divergence of magnetic fields 
admit the simplification
\begin{equation}                                                                 \label{eq:simpleAp}
\bm{A}(\bm{r}_1)\cdot\bm{p}_1=(\bm{r}_1\times\bm{p}_1)\cdot\bm{\tilde{B}}(\bm{r}_1)
=\bm{L}_1\cdot\bm{\tilde{B}}(\bm{r}_1) \quad , 
\end{equation}
where we exemplarily defined the angular momentum of particle 1, $\bm{L}_1=\bm{r}_1\times\bm{p}_1$.

Since the interaction potential depends only on the distance of the two particles, 
it is natural to introduce relative and c.m. coordinates, 
$\bm r_1=\bm R+(M_2/M)\bm r$ and 
$\bm r_2=\bm R-(M_1/M)\bm r$
with the total mass $M=M_1+M_2$.
If no external field was present, the new coordinates would decouple 
the internal degrees of freedom from the external c.m.\ ones. 
Yet even a homogeneous magnetic field 
couples the relative and the c.m.~motion \cite{dippel,peter:pla}. 
For neutral systems in static homogeneous magnetic fields, however, 
a so-called `pseudoseparation' can be performed providing us with an effective Hamiltonian for the relative motion, 
that depends on the c.m. motion only parametrically via the eigenvalues of the pseudomomentum
\cite{avron,peter:cpl208,peter:ctqc,dippel} which is associated with the c.m.\ motion.
Such a procedure is not available in the present case of a more general inhomogeneous field.
In the new coordinate system the Hamiltonian (\ref{eq:Hinit}) becomes
\begin{multline}
  H= H_0 
  + \bm{L}_1 \bm{\tilde{B}}(\bm{R}+\frac{M_2}{M}\bm{r})
  - \bm{L}_2 \bm{\tilde{B}}(\bm{R}-\frac{M_1}{M}\bm{r}) \\
  - \bm{\mu}_1\bm{B}(\bm R+\frac{M_2}{M}\bm r)
  - \bm{\mu}_2\bm{B}(\bm R-\frac{M_1}{M}\bm r) 
  + \mathcal O(\bm A^2)\ ,
\end{multline}
where the angular momenta of the particles read
\begin{align}
  \bm{L}_{1} = ({M_{1}}/{M})\bm{L}_R 
  + ({M_{2}}/{M})\bm{L}_r 
  + \bm{R}\times \bm{p} 
  + ({m}/{M}) \bm{r}\times\bm{P} \nonumber \\
  \bm{L}_{2} = ({M_{2}}/{M})\bm{L}_R 
  + ({M_{1}}/{M})\bm{L}_r 
  - \bm{R}\times \bm{p} 
  - ({m}/{M}) \bm{r}\times\bm{P} \nonumber 
\end{align}
(see also Ref.\ \cite{igor:pra72}), and the terms that do not depend on the field are summarized to 
$H_0=\frac{\bm{p}^2}{2m}+\frac{\bm{P}^2}{2M}+V(\bm{r})$. 
Here, $\bm{L}_{\bm{r}}=\bm{r}\times\bm{p}$, $\bm{L}_{\bm{R}}=\bm{R}\times\bm{P}$, and the reduced mass $m=M_1 M_2/M$ have been introduced. 

To simplify the Hamiltonian we apply a unitary transformation 
that eliminates c.m.~momentum dependent coupling terms generated by the homogeneous field component
\begin{equation}                                                                  \label{eq:U}
  U=\exp\left\{\frac{i}{2}\,\bm{B}_c\times \bm{r}\cdot\bm{R}\right\} \; .
\end{equation}
$H_0$ transforms as follows 
\[  U^{\dagger}H_0 U= 
  H_0+\frac{1}{2}\bm{B}_c \left(-\frac{1}{m}\bm{R}\times \bm{p}+\frac{1}{M}\bm{r}\times\bm{P}\right)
  +\mathcal{O}(\bm{B}_c^2).\]
The transformation of the remaining terms generates exclusively additional terms, that are quadratic
with respect to the magnetic field. 
Exploiting now the fact that the mass of the ionic core is much larger than the mass of the valence electron, 
we only keep magnetic field dependent terms of the order of the inverse light mass $1/M_1$ 
(which becomes $1$ in atomic units).
We arrive at the Hamiltonian 
\begin{multline}                                                                  \label{eq:UHU}
  U^\dagger H U     = {\bm{p}^2}/{2} + U^{\dagger}V(\bm{r})U + {\bm{P}^2}/{2M} 
   + {1}/{2}\;\bm{L}_{\bm r}\cdot\bm{B}_c \\
   + \bm{A}_l(\bm{R}+\bm{r})\cdot\bm{p}                             
   +(\bm{L}_{\bm r} +\bm{R}\times\bm{p})\cdot\bm{\tilde{B}}_{n}(\bm R + \bm r) \\
   - \bm{\mu}_1\cdot\bm B(\bm R + \bm r)-\bm{\mu}_2\cdot \bm{B}(\bm R) \; .    
\end{multline}

The diamagnetic terms which are proportional to $\bm{A}^2$ 
(and herewith proportional to $\bm{B}^2$, see Eq.~(\ref{eq:simpleAp}))
have been neglected. 
Due to the unitary transformation $U$, $\bm R$-dependent terms 
that are quadratic in the Ioffe field strength $B$ do not occur 
and only an electronic term $B^2 (x^2+y^2)/8$ remains 
whose typical energy contribution amounts to $B^2 n^4/8\approx 10^{5} B^2$ for $n=30$. 
Besides we obtain a term quadratic in the field gradient $G$. 
The term quadratic in the Ioffe field is negligible 
in comparison with the dominant shift due to the linear Zeeman term 
as long as $B$ is significantly smaller than $10^4$~Gauss which is guaranteed in our case. 
Moreover, the c.~m.\ coordinate dependance of this diamagnetic term 
is much weaker than the c.~m.\ coordinate dependance of the terms linear in the field gradient. 
The term quadratic in the field gradient can be neglected in comparison with the corresponding linear term.
Up to now we did not use the explicit form of the Ioffe-Pritchard field configuration.
(In anticipation of the special field configuration we leave the term containing $\bm A_l$ in its original form.)

\subsection{\label{ch:ip}Ioffe-Pritchard Field Configuration}

Two widely spread magnetic field configurations that exhibit a local field minimum and serve
as key ingredients for the trapping of weak-field seeking atoms
are the 3D quadrupole and the Ioffe-Pritchard configuration.
The Ioffe-Pritchard configuration resolves the problem of particle loss due to spin flip
by means of an additional constant magnetic field.
A macroscopic realization uses four parallel current carrying Ioffe bars which generate the quadrupole field.
Encompassing Helmholtz coils create the additional constant field. There are many alternative layouts, 
the field of a clover-leaf trap for example features the same expansion around the origin \cite{cloverleaf}. 
On a microscopic scale the Ioffe-Pritchard trap has been implemented on atom chips by a $Z$-shaped wire \cite{folman}.  

The vector potential and the magnetic field read
\begin{gather}  
  \label{eq:overallpotential}
  \bm{A}=\underbrace{ \frac{B}{2} \left( \begin{array}{ccc} -y \\ x \\ 0 \end{array} \right)}_{=\bm{A}_c}
  +\underbrace{G\left( \begin{array}{ccc} 0 \\ 0 \\ x y \end{array} \right)}_{=\bm{A}_l}
  +\bm{A}_q \; , \\
  \label{eq:overallfield}
  \bm{B}= \underbrace{B \left( \begin{array}{ccc} 0 \\ 0 \\ 1 \end{array} \right)}_{=\bm{B}_c}
  +\underbrace{G \left( \begin{array}{ccc} x \\ -y \\ 0 \end{array} \right)}_{=\bm{B}_l}
  +\bm{B}_q \; .
\end{gather}
where 
$\bm{A}_q=\frac{Q}{4}(x^2+y^2-4 z^2) (-y \mathbf{e}_x + x \mathbf{e}_y)$ and
$\bm{B}_q=Q (2xz\mathbf{e}_x + 2yz\mathbf{e}_y +  (x^2+y^2-2 z^2)\mathbf{e}_z)$.
$\bm{B}_c$ is the constant field created by the Helmholtz coils with $B$ being the Ioffe field strength. 
$\bm{B}_l$ originates from the Ioffe bars and depends on the field gradient $G$.
$\bm{B}_q$ designates the quadratic 
term generated by the Helmholtz coils whose magnitude, compared to the first Helmholtz term, can be varied by changing the geometry of the trap, 
$Q= B\cdot \frac{3}{2}(R^2-4D^2)/(R^2+D^2)^{2}$, 
where $R$ is the radius of the Helmholtz coils, and $2D$ is their distance from each other.

If we now insert the special Ioffe-Pritchard field configuration using Eqs.~(\ref{eq:overallpotential},\ref{eq:overallfield}) 
into the transformed Hamiltonian (\ref{eq:UHU}) we obtain
\begin{align}                                                                                 \label{eq:HIP}
  &H_{IP} = H_A +{\bm{P}^2}/{2M} \nonumber 
+{BL_z}/{2}  +G(x+X)(y+Y)p_z \nonumber\\
  &+Q/4\big[\big((x+X)p_y-(y+Y)p_x\big)     \nonumber\\  
  &\cdot\big((x+X)^2+(y+Y-2(z+Z))(y+Y+2(z+Z))\big)\big]\nonumber\\
  &- \bm{\mu}_1 \bm B(\bm R + \bm r)-\bm{\mu}_2\bm{B}(\bm R) \; ,
\end{align}
where $H_A = {\bm p^2}/{2}-{1}/{r}$ is the operator for a field free atom.
The well known Zeeman term ${BL_z}/{2}$ comes from the uniform Ioffe field generated by the Helmholtz coils.
The following term, involving the field gradient $G$, arises from the linear field generated by the Ioffe bars and
couples the relative and c.m.\ dynamics.
The part in the squared brackets originates from the quadratic term, again created by the coils.
It is the only one that depends on the $Z$ coordinate, 
we will see below that its contribution is negligible under certain conditions. 
The last term couples the spin of particle two to the magnetic field. 
Since the electronic spins of closed shells combine to zero, the spin of particle two is the \emph{nuclear} spin only.
Even though $\bm \mu_2\bm B$ scales with ${1}/{M_2}$, we will still keep the term.
Being the only one containing the nuclear spin it is essential for a proper symmetry analysis.

\subsection{\label{ch:SSS}Symmetries, Scaling and the Approximation of a Single $n$-Manifold}

Our Hamiltonian is invariant under a number of symmetry transformations $U_S$ 
that are composed of the elementary operations listed in Tab.~\ref{t:sym}.
The parity operations $P_j$, $j \in \{x,y,z\}$, are defined by their action on the spatial laboratory coordinates of the particles 
which translates one-to-one to c.m. and relative coordinates.
In order to exchange the $x$ and $y$ components of the electronic spin we introduce the operator
\[S_{xy}=  \begin{pmatrix} -i&0\\ 0&1 \end{pmatrix} \; , \]
where $S_{xy} S_{xy}^{*}=1$. 
$T$ represents the conventional time reversal operator for spinless particles which, 
in the spatial representation,
corresponds to complex conjugation.
Our unitary symmetries are
\begin{subequations}\label{eq:sym}
  \begin{gather}
    P_x P_y \hat S_z \hat \Sigma_z                                    \label{eq:symI} \\
    P_y P_z I_{xy} S_{xy} \Sigma_{xy}                                 \label{eq:symII} \\
    P_x P_z I_{xy} S^{*}_{xy} \Sigma^{*}_{xy} \; .                    \label{eq:symIII}
  \end{gather}
\end{subequations}
The Hamiltonian is also left invariant under the antiunitary symmetry transformation
\begin{equation}
T P_y  .                                                          \label{eq:symA} \\
\end{equation}
By consecutively applying the latter operator and the unitary operators (\ref{eq:symI}), (\ref{eq:symII}) and (\ref{eq:symIII})
it is possible to create further antiunitary symmetries:
\begin{subequations}\label{eq:symAall}
  \begin{gather}
    T P_x \hat S_z \hat \Sigma_z                                        \label{eq:symAI} \\
    T P_z I_{xy} S_{xy} \Sigma_{xy}                                     \label{eq:symAII} \\
    T P_x P_y P_z I_{xy} S^{*}_{xy} \Sigma^{*}_{xy} .                   \label{eq:symAIII}
  \end{gather}
\end{subequations}
Paying regard to the fact that $S_{xy}^2=-\hat S_z$ and $\Sigma_{xy}^2=-\hat \Sigma_z$
and that $T$ neither commutes with $\hat S_y$ nor with $S_{xy}$ and $\Sigma_{xy}$,
one finds that the operators (\ref{eq:symI}-\ref{eq:symAIII}) form a symmetry group.

\begin{table}[tb] 
  \begin{center}
    \begin{tabular}{lll}
      \hline \\[-12pt]
      operator &            & operation\\
      \hline \hline \\[-9pt]
      $P_{x}$ & $x$ parity & $x\rightarrow -x$, $X\rightarrow -X$ \\
      $\hat S_x$ & electronic spin $x$ op. & $S_y\rightarrow -S_y$, $S_z\rightarrow -S_z$ \\
      $\hat \Sigma_x$ & nuclear spin $x$ op. & $\Sigma_y\rightarrow -\Sigma_y$, $\Sigma_z\rightarrow -\Sigma_z$ \\
      $I_{xy}$ & coordinate exchange & $x\leftrightarrow y$, $X\leftrightarrow Y$ \\
      $S_{xy}$ & el.\ spin component exc.& $S_x\rightarrow -S_y$, $S_y\rightarrow S_x$ \\
      $\Sigma_{xy}$ & nuclear spin comp. exc.&  $\Sigma_x\rightarrow -\Sigma_y$, $\Sigma_y\rightarrow \Sigma_x$ \\
      $T$ & conventional time reversal & $A\rightarrow A^{*}$ \\
      \hline \hline \\[-25pt]
    \end{tabular}
  \end{center}
  \caption[Symmetry operation nomenclature]{\label{t:sym} Symmetry operation nomenclature. 
    $P_j$, $\hat S_j$, and $\hat \Sigma_j$ are exemplified by $j=x$, but hold of course also for $j=y,z$.}
\end{table}

If no Ioffe field is present ($B=0$), eight additional symmetries can be found leaving the Hamiltonian invariant.
For an effective one particle approach (and the corresponding one particle symmetries) 
this was discussed in Ref.~\cite{igor:pra70:4}.


As indicated before, the quadratic magnetic field term is small 
and can be tuned by changing the trap geometry. 
It can provide a longitudinal confinement which may be treated by perturbative methods. 
In the case of negligible quadratic field $\bm B_q$, which we assume in the following, 
the term in the squared brackets of the Hamiltonian (\ref{eq:HIP}) drops out 
and the $Z$ coordinate is cyclic. 
The corresponding conjugated momentum $P_z$ is consequently conserved 
and the longitudinal motion is integrated by simply employing plane waves 
$| k_Z\rangle = \exp\{i Z k_Z\}$.
The constraints for this approximation to be valid can be obtained by comparing 
the above-mentioned term in squared brackets with the Zeeman term, $BL_z/2$.
Estimating $\langle x\rangle \approx n^2$,
$\langle xp_y\rangle \approx \langle yp_x\rangle \approx n$, and
using $|Q|\lessapprox B/(D^2+R^2)$
we find
\begin{align}
  D^2+R^2 &\gg n^4 \quad\quad  \textrm{and} \label{eq:constraintA}\\
  \sqrt[3]{n(D^2+R^2)} &\gg  X,Y \; ,\label{eq:constraintB}
\end{align}
where $D$ and $R$ characterize the trap geometry.
Eqs.~(\ref{eq:constraintA},\ref{eq:constraintB}) are easily fulfilled. 
We are therefore left with the Hamiltonian 
\begin{equation}                                                                     \label{eq:Hnotscaled}
  H = H_A + (P_x^2+P_y^2)/{2} + H_e \; ,
\end{equation}
where the electronic Hamiltonian reads 
\begin{equation}                                                                     \label{eq:Henotscaled}
  H_{e} = {BL_z}/{2}  +G(x+X)(y+Y)p_z - \bm{\mu}_1 \bm B(\bm R + \bm r) \; .
\end{equation}

For all laboratory fields one finds the magnetic field strength $B$ and the magnetic field gradient $G$ 
to be a lot smaller than 1. 
Our Hamiltonian (\ref{eq:HIP}) is thus dominated by $H_A$.
The energies of the field free spectrum $E_A^n= -1/2 n^2$ are $n^2$-fold degenerate.
We can assume the Ioffe-Pritchard field not to couple adjacent $n$-manifolds 
as long as $|{E_A^{n}-E_A^{n\pm 1}}|/{E_{Zee}} \gg 1$. 
The resulting constraints $B \ll n^{-4}$,  $G\ll n^{-6}$ and $GR\ll n^{-4}$ 
yield $B \ll 2900$~G, $G\ll 6\cdot 10^{6}$~T/m for $n=30$ and $R\ll 2.9$~mm 
if we additionally assume the field gradient $G$ to be as large as $100$~T/m. 
In our parameter regime each $n$-manifold can therefore be considered separately.
We thus project the full Hamiltonian on 
the hydrogenic eigenfunctions $| \alpha \rangle = | n, l, m_l, m_s \rangle$, $H_A | \alpha \rangle = E_A^n | \alpha \rangle $,
with fixed principal quantum number $n$, that cover an entire n-manifold.
$l$ denotes the orbital angular momentum quantum number, 
$m_l$ the one of its $z$ component $L_z$
and $m_s$ stands for the quantum number of the electronic spin.

Working in a single $n$-manifold 
we can reformulate the term in the Hamiltonian (\ref{eq:Hnotscaled})
involving the field gradient $G$ into a more compact form.
We first consider the commutator
$[yz, H_A] = [yz, \bm p^2]/2 = i (yp_z + zp_y)$. 
This yields 
\begin{equation}
  \langle\alpha | yp_z |\alpha^{\prime}\rangle + \langle\alpha | zp_y |\alpha^{\prime}\rangle 
  = -i \langle\alpha | [yz,H_A] |\alpha^{\prime}\rangle = 0 \; ,
\end{equation}
since $|\alpha\rangle$ and $|\alpha^\prime\rangle$ are eigenkets 
to the same eigenvalue $E_n$.
Establishing the relation to the orbital angular momentum operator 
via $yp_z= L_x + zp_y$ results in
\begin{equation}
  (\langle\alpha | yp_z |\alpha^{\prime}\rangle) = \frac{1}{2} (\langle\alpha | L_x |\alpha^{\prime}\rangle) \; . 
\end{equation}
The same procedure can be applied to $xp_z$ leading to 
\begin{equation}
  (\langle\alpha | xp_z |\alpha^{\prime}\rangle) = -\frac{1}{2} (\langle\alpha | L_y |\alpha^{\prime}\rangle) \; . 
\end{equation}
Furthermore $\langle \alpha | XYp_z | \alpha' \rangle = 0$ 
since $p_z \sim [H_A,z]$, and eventually we can write
\begin{equation}
  G(x+X)(y+Y)p_z = G(xyp_z+XL_x/2-YL_y/2)\; ,
\end{equation}
where we omitted the bracketing alphas, 
but keep in mind that the above identity holds in a single $n$-manifold only. 

In order to remove the separate dependencies on the field parameters $B$, $G$, and on the mass $M$ from the coupling terms, 
we introduce scaled c.m.~coordinates, 
$\mathbf{R}\rightarrow \gamma^{-\frac{1}{3}} \mathbf{R}$, with $\gamma=G M$, 
and simultaneously we introduce the energy unit $\epsilon=\gamma^\frac{2}{3}/M$. 
Introducing the effective magnetic field
\begin{equation}
  \bm{G}(X,Y) = \left( \begin{array}{ccc} X \\ -Y \\ \zeta \end{array} \right) \; ,
  \quad \zeta =BM\gamma^{-\frac{2}{3}}\; ,
\end{equation}
and omitting the constant energy offset $E_A^n$, 
the Hamiltonian can be given the advantageous form
\begin{equation}                                                                                        \label{eq:Hwork}
  \mathcal H= \frac{P_x^2+P_y^2}{2}
  + \bm{\mu}\cdot\bm{G}(X,Y) 
  + \gamma^{\frac{1}{3}} (xyp_z+x S_x-y S_y).
\end{equation}
The first term is the c.m.~kinetic energy. 
$\bm \mu$ is the $2n^2$-dimensional matrix representation of the total magnetic moment of the electron,
$\frac{1}{2}(\bm L_{\bm r} +2\bm S)$, 
and the second term in (\ref{eq:Hwork}) describes its coupling to the effective magnetic field $\bm{G}$. 
The latter results from the original field~$\bm B_c+\bm B_l$ in Eq.~(\ref{eq:overallfield})
taking into account the corresponding coordinate and energy scaling factors. 
$S_i$ are the components of the electronic spin, $\bm S=-\bm\mu_1$. 
The nuclear spin term $-\bm{\mu}_2\cdot\bm{B}(\bm R)$ has been omitted since it is several orders of magnitude smaller than the electronic one.

\section{\label{s:aa}Adiabatic Approach}

The large difference of the particles' masses and velocities in our two body system makes it plausible 
to adiabatically separate the electronic and the c.m.~motion.
The corresponding time scales differ substantially even for large principal quantum numbers $n$. 
However, due to the enormous level density in case of Rydberg atoms it is \emph{a priori} unclear 
whether isolated energy surfaces might exist or whether, as one might naturally assume, 
non-adiabatic couplings are ubiquitous and therefore an adiabatic approach might invalidate itself.
The procedure is reminiscent of the Born-Oppenheimer ansatz in molecular systems and is based on the idea that 
the slow change of the heavy particle's position  allows the electron to adapt instantaneously to the inhomogeneous field.
The electronic energy of the system can thus be considered as a function of the position of the heavy particle.

The adiabatic approximation is introduced by subtracting the transversal c.m.~kinetic energy, $\mathcal T={(P_x^2+P_y^2)}/{2}$, 
from the total Hamiltonian (\ref{eq:Hwork}).
The remaining electronic Hamiltonian for fixed center of mass reads 
\begin{equation}                                                                                       \label{eq:BOHe}
  \mathcal H_e=  \bm{\mu}\cdot\bm{G}(X,Y)   + \gamma^{\frac{1}{3}} (xyp_z+x S_x-y S_y).
\end{equation}
The electronic wave function $\varphi_\kappa$ depends parametrically on $\bm R$ 
and the total atomic wavefunction can be written as 
\begin{equation}
  |\Psi(\bm r,\bm R)\rangle =
  |\varphi_\kappa(\bm r;\bm R)\rangle
  \otimes |\psi_\nu(\bm R)\rangle \; ,
\end{equation}
where $|\psi_\nu(\bm R)\rangle$ is the center of mass wave function. 
The internal problem posed by the stationary, electronic Schr\"odinger equation
\begin{equation}                                                                                         \label{eq:internal}
  \mathcal H_e \; |\varphi_\kappa(\bm r;\bm R)\rangle = E_\kappa(X,Y)\; |\varphi_\kappa(\bm r;\bm R)\rangle
\end{equation}
is solved for the adiabatic electronic potential energy surfaces $E_\kappa(X,Y)$, 
that serve as a potential for the c.~m.~dynamics.
Within this approximation, the equation of motion for the center of mass wave function reads
\begin{equation}                                                                                         \label{eq:BOcm}
  \left( \mathcal T + E_\kappa(X,Y) \right) \; |\psi_\nu(\bm R)\rangle = \epsilon_{\nu} \; |\psi_\nu(\bm R)\rangle \; .
\end{equation}

The spatially dependent transformation $\mathcal U(X,Y)$, 
that diagonalizes the matrix representation $\mathcal{H}_e$ of the electronic Hamiltonian, 
is composed of the vector representations of the electronic eigenfunctions, 
$\bm{\mathcal{U}}_\kappa = \left( \mathcal{U}_{\kappa\alpha} \right) = \left( \langle \alpha |\varphi_\kappa(\bm r;\bm R)\rangle \right)$.
Since $\mathcal{U}$ depends on the c.m.~coordinates, the transformed kinetic energy involves 
non-adiabatic couplings \nolinebreak $\Delta \mathcal T$
\begin{equation}
  \mathcal{U}^\dagger \mathcal H \mathcal{U} = \mathcal{U}^\dagger \mathcal H_e \mathcal{U} + \mathcal{U}^\dagger \mathcal T \mathcal{U} = E_\kappa(X,Y) + \mathcal T + \Delta \mathcal T
\end{equation}
that have been neglected in the adiabatic approximation of Eq.~(\ref{eq:BOcm}), 
\begin{multline}                                                                 \label{eq:DeltaT}
  \Delta \mathcal T = -1/2 \cdot\big(
  \mathcal{U}^\dagger (\partial^2_X \mathcal{U})
  + \mathcal{U}^\dagger (\partial^2_Y \mathcal{U}) \\
  + 2 \mathcal{U}^\dagger (\partial_X \mathcal{U}) \partial_X       
  + 2 \mathcal{U}^\dagger (\partial_Y \mathcal{U})\partial_Y \big) \; .
\end{multline}
They can be calculated explicitly as soon as the electronic adiabatic eigenfunctions 
have been computed.
Non-adiabatic contributions can be neglected if the conditions 
\begin{gather}                                                                                                  \label{eq:neglectdeltaT1}
  \rvert \frac{\langle\varphi_{\kappa\prime} | (\partial_X \mathcal H)  |\varphi_\kappa\rangle }{E_{\kappa\prime}-E_\kappa} \rvert \ll 1  
  \; , \quad
  \rvert \frac{\langle\varphi_{\kappa\prime} | (\partial_Y \mathcal H)  |\varphi_\kappa\rangle }{E_{\kappa\prime}-E_\kappa} \rvert \ll 1 
  \; ,\\ \label{eq:neglectdeltaT2}
  \rvert \frac{\langle\varphi_{\kappa\prime} | (\partial^2_X \mathcal H)  |\varphi_\kappa\rangle }{E_{\kappa\prime}-E_\kappa} \rvert \ll 1  
  \; , \quad
  \rvert \frac{\langle\varphi_{\kappa\prime} | (\partial^2_Y \mathcal H)  |\varphi_\kappa\rangle }{E_{\kappa\prime}-E_\kappa} \rvert \ll 1 
  \phantom{\; ,}
\end{gather}
are fulfilled \cite{igor:pra72}. 
The energy denominator in (\ref{eq:neglectdeltaT1}) and (\ref{eq:neglectdeltaT2}) indicates that
one can expect non-adiabatic couplings to become relevant between the adiabatic energy surfaces
 when they come very close in energy, i.e.\ in the vicinity of avoided crossings.

Recalling the results of the symmetry analysis, 
it can be demonstrated that the energy surfaces $E_\kappa$, exhibit three mirror symmetries.
Within the adiabatic approximation,  
$X$ and $Y$ are parameters in the electronic Schr\"odinger equation. 
Symmetry operations applied to the electronic Hamiltonian 
thereby merely act onto the electronic subspace.  
If we apply the corresponding restricted symmetry operation $U_{P} = P_x P_y \hat S_z \hat \Sigma_z$ (\ref{eq:symI}),
that was already shown to leave the full Ioffe-Pritchard Hamiltonian (\ref{eq:HIP}) invariant,
to the electronic Hamiltonian $H_e$ (\ref{eq:Henotscaled}), we find 
\begin{equation}
  U_{P}^\dagger  H_e(\bm r; X,Y)  U_{P} =  H_e(\bm r; -X,-Y) \; .
\end{equation}
Since unitarily equivalent observables, $A$ and $U^\dagger A U$,
possess the same eigenvalue spectrum, 
we find the energy surfaces to be inversion symmetric with respect to the origin in the $X$-$Y$ plane.
The symmetry operator $U_{Y}=T P_y$, 
and the operator that is composed of $U_{Y}$ and $U_{P}$, namely
$U_{X}=T P_x \hat S_z \hat \Sigma_z$ 
(see (\ref{eq:symA}) and (\ref{eq:symAI})),
mirror the energy surfaces at the axes, 
\begin{align}
  U_{Y}^\dagger H_e(\bm r; X,Y) U_{Y} &=  H_e(\bm r; X,-Y) \; ,\\
  U_{X}^\dagger H_e(\bm r; X,Y) U_{X} &=  H_e(\bm r; -X,Y) \; .
\end{align}

The electronic problem (\ref{eq:internal}), with the core fixed at an arbitrary position, is three-dimensional. 
No symmetry arguments can be exploited to reduce the dimensionality of the problem. 
In order to solve it, we employ the variational method, which maps the stationary Schr\"odinger equation
onto an ordinary algebraic eigenvalue problem. 
Since the matrix representation of the electronic Hamiltonians is sparsely occupied, 
an Arnoldi decomposition is used.
Both, this decomposition and the surfaces' mirror symmetries, help to reduce the computational cost
of solving the electronic Schr\"odinger equation.

\section{\label{s:ees}Electronic potential energy surfaces}
In this section the properties of the electronic adiabatic energy surfaces are analyzed 
for different regimes of Ioffe field strengths and field gradients.
These two parameters can be used 
to shape the potential in which the center of mass dynamics takes place.
To understand how this takes place, 
we inspect the electronic Hamiltonian to
unravel the influence of the individual terms for different parameter regimes.

The characteristic length scale of the center of mass dynamics 
is of the order of one in scaled atomic units. 
It is therefore adequate to compare the magnitudes of the different parts of the electronic Hamiltonian (\ref{eq:BOHe})
in order to estimate their impact on the center of mass motion, putting $X$ and $Y$ equal to one. 
The first part, $\bm{\mu}\cdot\bm{G}(X,Y)$, 
consists of the coupling terms $X(\frac{1}{2}L_x+S_x) - Y(\frac{1}{2}L_y+S_y)$, 
that are then of the order of $\langle L_i \rangle \approx n$ for high angular momentum states, 
and of the Zeeman term $\zeta(\frac{1}{2}L_z+S_z)$, which can be as large as $\zeta n$. 
The second part, $\gamma^{{1}/{3}}(xyp_z+x S_x-y S_y)$, 
is quadratic in the relative coordinates 
which makes it particularly important for high principal quantum numbers $n$.
If we consider the expectation values of the relative coordinates to be of the order of $n^2$, and 
$\langle yp_z \rangle \approx \langle L_x \rangle \approx n$, 
the overall magnitude can be estimated to $\gamma^{{1}/{3}} n^3$.
In a nutshell, we have for the mentioned three terms the following relative orders of magnitude, 
\begin{equation}                                                                                                       \label{eq:factors}
  1  \; , \quad  \zeta \quad \textrm{and} \quad  \gamma^{\frac{1}{3}} n^2 \; .
\end{equation}
Due to the special form of the electronic Hamiltonian, 
changing the magnetic field parameters~$B$ and~$G$ while keeping their ratio~$\zeta/ \gamma^{{1}/{3}}=B/G$ (and $n$) constant 
results in a mere scaling of the c.m.~coordinates. 
We provide typical examples for values of the quantities (\ref{eq:factors}) in table \ref{t:parameters}.

\begin{table}[tbp]
\centering
\begin{small}
\begin{tabular}{lr|ccccccc|}
\\[-5pt]
\multicolumn{3}{r}{G [T/m]\quad\; \textbf{0.01$\phantom{||}$}} & \textbf{0.1} & \textbf{1} & \textbf{10} & \textbf{100} & \textbf{1000} & \multicolumn{1}{l}{\textbf{10000}} \\ 
\hline\\[-12pt]
\multicolumn{1}{r}{\textbf{}} & \multicolumn{1}{r}{n\, } & \multicolumn{1}{l}{\textbf{}} & \multicolumn{1}{l}{\textbf{}} & \multicolumn{1}{l}{\textbf{}} & \multicolumn{1}{l}{\textbf{}} & \multicolumn{1}{l}{\textbf{}} \\ \cline{3-9}
\multicolumn{1}{r}{\textbf{}} & & \multicolumn{1}{l}{\textbf{}} & \multicolumn{1}{l}{\textbf{}} & \multicolumn{1}{l}{\textbf{}} & \multicolumn{1}{l}{\textbf{}} & \multicolumn{1}{l}{\textbf{}} &  &\\ [-12pt]
{\colorbox{mygrey}{$\gamma^{\frac{1}{3}}n^2$}} & \textbf{3} & 0.001 & 0.001 & 0.003 & 0.006 & 0.014 & 0.030 & 0.064 \\ 
 & \textbf{10} & 0.007 & 0.015 & 0.033 & 0.071 & 0.153 & 0.329 & 0.709 \\ 
 & \textbf{30} & 0.064 & 0.138 & 0.296 & 0.638 & 1.375 & 2.963 & 6.383 \\ 
 & \textbf{50} & 0.177 & 0.382 & 0.823 & 1.773 & 3.820 & 8.229 & 17.729 \\ 
 & \textbf{80} & 0.454 & 0.978 & 2.107 & 4.539 & 9.778 & 21.067 & 45.387  \\ \cline{3-9}
\\[-10pt]
\multicolumn{3}{c}{B [Gauss]\;\;} & \multicolumn{1}{l}{} & \multicolumn{1}{l}{} & \multicolumn{1}{l}{} & \multicolumn{1}{l}{} & \multicolumn{1}{l}{} \\ \cline{3-9}
\multicolumn{1}{r}{\textbf{}} & & \multicolumn{1}{l}{\textbf{}} & \multicolumn{1}{l}{\textbf{}} & \multicolumn{1}{l}{\textbf{}} & \multicolumn{1}{l}{\textbf{}} & \multicolumn{1}{l}{\textbf{}} &  &\\ [-12pt]
{\colorbox{mygrey}{$\zeta$}} & \textbf{0.01} & 134.0 & 28.87 & 6.220 & 1.340 & 0.289 & 0.062 & 0.013 \\ 
 & \textbf{0.1} & 1340 & 288.7 & 62.20 & 13.40 & 2.887 & 0.622 & 0.134 \\ 
 & \textbf{1} & 13402 & 2887 & 622.0 & 134.0 & 28.87 & 6.220 & 1.340 \\ 
 & \textbf{10} & 134015 & 28873 & 6220 & 1340 & 288.7 & 62.20 & 13.40 \\ \cline{3-9}
\\[-8pt]
\hline
\end{tabular}
\end{small}
\caption[Parameters]{Explicit values for $\gamma^{{1}/{3}} n^2=(GM)^{{1}/{3}} n^2$ 
  and $\zeta=BM^{{1}/{3}}G^{-{2}/{3}}$ for $^{87}$Rb in atomic units. 
  The first block lists  $\gamma^{{1}/{3}} n^2$ for different values of the field gradient $G$ and for different principal quantum numbers~$n$. 
  The second block lists~$\zeta$ for different field gradients and for different field strengths~$B$.
}
\label{t:parameters}
\end{table}

\subsection{\label{ch:regulatingcapacity}Regulating Capacity of the Ioffe Field}

To understand the impact of the Ioffe field strength $B$ on the adiabatic energy surfaces, 
we isolate its effect by suppressing other influences.
This can be done by choosing a relatively low field gradient $G$ and/or a small principal quantum number $n$ (see Tab.~\ref{t:parameters}). 
The factor $\gamma^{\frac{1}{3}} n^2$ becomes small, and the last term in Eq.~(\ref{eq:BOHe}) will hardly provide any contribution.
Within this regime, that we focus on in this subsection, 
approximate analytical expressions for the electronic adiabatic energy surfaces can be derived. 
We diagonalize the approximate electronic Hamiltonian 
\begin{equation}                                                                                \label{eq:Hetilde}
  \tilde H_e= \frac{1}{2} \bm{G} \, (\bm L + 2\bm S) \; .
\end{equation}
by applying the spatially dependent unitary transformation
\begin{equation}                                                                                                         \label{eq:diagtrafo}
  U_D(X,Y)=  e^{i\phi(L_z+S_z)}  e^{i\beta(L_y+S_y)} \; , 
\end{equation}
with $\phi=\arctan\frac{Y}{X}$,
$\cos\beta=\gamma^{-\frac{2}{3}}M_2 B|\bm{G}(X,Y)|^{-1}$ 
and $\sin\beta=-\sqrt{X^2+Y^2}|\bm{G}(X,Y)|^{-1}$. 
This yields 
\begin{equation}
  U^\dagger_D \tilde H_e U_D = \frac{1}{2} ( L_z + 2 S_z) |\bm G(X,Y)| \; 
\end{equation}
for the transformed approximate electronic Hamiltonian. 
The spatially dependent transformation~$U_D$ locally rotates the magnetic moment of the electron, 
which includes its spin and its angular momentum, 
such that it is parallel to the local direction of the magnetic field. 
The operators~$L_z$ and~$S_z$ are not identical to the ones before having applied the transformation~(\ref{eq:diagtrafo}), 
they are rather related to the local quantization axis defined by the local magnetic field direction~\cite{igor:prl}.

The adiabatic potential surfaces evaluate to 
\begin{align}                                                                                                  \label{eq:Ekappasmallgradient}
  E_\kappa(X,Y) &= \frac{1}{2} (m_l + 2m_s) |\bm G(X,Y)| \nonumber\\
  &= \frac{1}{2} (m_l + 2m_s) \sqrt{X^2+Y^2+\zeta^2} \; .
\end{align}
The possible combinations of $m_l$ and $m_s$ yield $2n+1$ energy surfaces. 
The surfaces highest and lowest in energy correspond to circular states, ($|m_l|=l_{max}=n-1$, $m_l+2m_s=\pm n$), 
and they are the only non-degenerate ones. 
For the other surfaces ($|m_l+2m_s| < n$), 
the multiplicity of $(m_l+2m_s)$, 
and with that the degree of degeneracy of the corresponding surfaces, 
is given by $2n - |m_l+2m_s+1| - |m_l+2m_s-1|$. 
Starting from the highest energy surface, 
the levels of degeneracy thus are 1, 2, 4, 6, \dots. 

The approximate surfaces $E_\kappa$ (\ref{eq:Ekappasmallgradient}) are rotationally symmetric around the $z$-axis. 
An expansion around this axis ($\rho = \sqrt{X^2+Y^2} \ll \zeta$) 
yields a harmonic potential, 
\begin{equation}                                                                                                        \label{eq:Eksmallrho}
  E_\kappa(\rho) \approx (\zeta + \frac{1}{2\zeta} \rho^2)\cdot\frac{1}{2} (m_l + 2m_s) \; ,
\end{equation}
while we find a linear behavior, 
\begin{equation}                                                                                                         \label{eq:Ekbigrho}
  E_\kappa(\rho) \approx \frac{\rho}{2}\cdot(m_l + 2m_s) \; ,
\end{equation} 
when the center of mass is far from the $z$-axis ($\rho \gg \zeta$). 

\begin{figure}[tbh!]
  \centering
  \includegraphics[width=8cm]{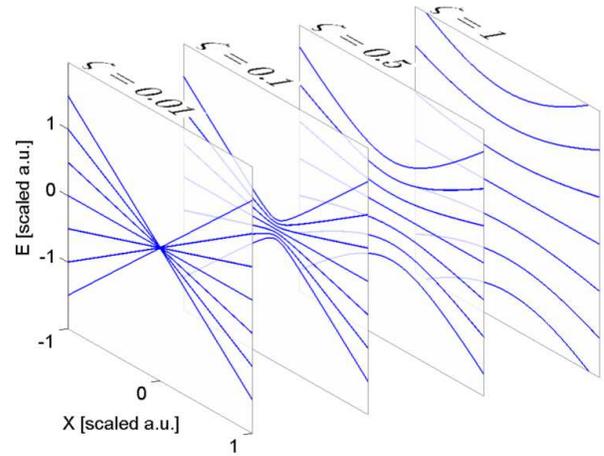}
  \caption[Sections for increasing Ioffe field, $n=3$.]{Sections along the $X$-axis through the electronic adiabatic energy surfaces 
    of an entire $n=3$ manifold. The field gradient is fixed at $G=1$ Tesla/m in order to suppress the influence 
    of the last term in $H_e$ (\ref{eq:BOHe}).
    From left to right, $\zeta=B M \gamma^{-{2}/{3}}$ increases due to an increasing Ioffe field.}                     \label{f:differentzetas}
\end{figure} 

For reasons of illustration we demonstrate the behavior of the adiabatic surfaces with increasing Ioffe field 
by means of a somewhat artificial example where other, previously neglected interactions might be more important. 
Fig.~\ref{f:differentzetas} shows sections through all the surfaces for $n=3$. 
This principal quantum number has been chosen in order to keep the sections simple while displaying the entire $n$-manifold. 
We employ $^{87}$Rb in this expository example although the electronic ground state of its outermost electron is $5$s. 
The sections have been calculated for the field gradient $G=1$~T/m and for different field strengths~$B$ 
using the total electronic Hamiltonian (\ref{eq:BOHe}). 
These parameters yield $\gamma^{{1}/{3}} n^2=0.003$, and values for~$\zeta$ ranging from $0.01$ to $1$. 
The surfaces in the different graphs of Fig.~\ref{f:differentzetas} indeed 
validate the approximate expression~(\ref{eq:Ekappasmallgradient}): 
We find $2n+1$ degenerate surfaces and 
the harmonic behavior for $|X| \ll \zeta$ gives way to a linear increase for $|X| \gg \zeta$. 
The energetic distances and lengths in the different graphs are comparable, 
since the scaling factor for the center of mass coordinates $\gamma=c_2M$ has not been changed. 
We can conclude that increasing the Ioffe field strength $B$ separates the surfaces from each other.

 \begin{figure}[tb!]
   \centering
     \includegraphics[width=7.5cm]{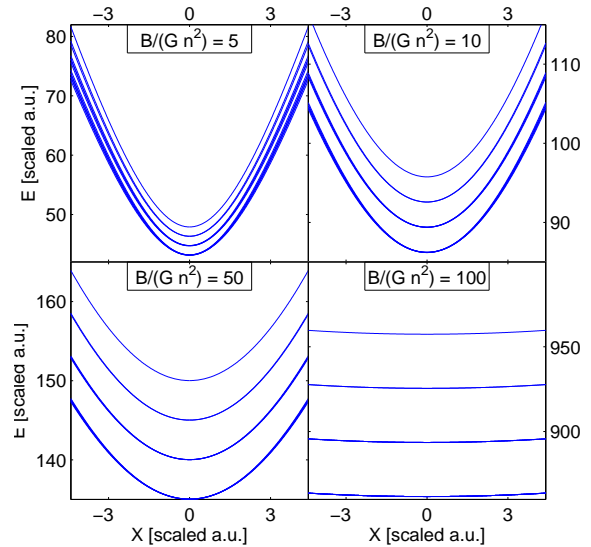} 
     \caption[Sections for increasing Ioffe field, $n=30$.]{Sections along the $X$-axis through the uppermost $21$ surfaces 
       of the $n=30$ manifold of $^{87}$Rb for increasing ratios $B/(G n^2)$. 
       The field gradient is fixed at $G=10$ T/m while the Ioffe field is increased from top left to bottom right. 
       ($B=24$ mG, $B=48$ mG, $B=0.24$ G, $B=0.48$ G). 
       For small ratios $B/(G n^2)$  the influence of the second term in (\ref{eq:BOHe}) is not completely suppressed 
       as can be seen from the lifted degeneracies in the upper subfigures.}                      \label{f:n30differentzetas}
 \end{figure} 

The data presented in Fig.~\ref{f:n30differentzetas} have been computed for the $n=30$ manifold. 
In order to keep the last term in (\ref{eq:BOHe}) small, the field gradient has been set to $G=0.1$~T/m 
($\rightarrow \gamma^{{1}/{3}} n^2=0.14$).
The uppermost $21$ energy surfaces are shown for different values of the magnetic field strength $B$.
Similar to the $n=3$ case, one can see the harmonic behavior around the origin. 
The surfaces' minimal distance becomes larger for increasing $\zeta$.  
Since $\zeta$ and $\gamma^{{1}/{3}} n^2$ are of the same order of magnitude in subfigure (a), 
the contribution of the last term in (\ref{eq:BOHe}),
that lifts the degeneracy of the curves, is visible.

The energetic distance of the approximate surfaces described by Eq.~(\ref{eq:Ekappasmallgradient}) 
increases with larger distances from the $Z$-axis, $\rho$, 
and with larger $\zeta$. 
The minimum energetic gap between two adjacent surfaces is at the origin and reads 
\begin{equation}                                                                                                            \label{eq:mindist}
  | E_\kappa(O) - E_{\kappa\pm 1}(O) | 
= \frac{B}{2} M\gamma^{-\frac{2}{3}} 
= \frac{\zeta}{2} \; . 
\end{equation}
The parameter $\zeta$ (an hence the field strength $B$) is the tool to control the energetic distance between the adiabatic surfaces.
Increasing $\zeta$, one can thus also minimize the non-adiabatic couplings $\Delta \mathcal T$ (\ref{eq:DeltaT}) 
discussed in Sect.~\ref{s:aa}, 
since they scale with the reciprocal energetic distance of the surfaces. 

   \begin{table}[bt]                                                                                         
   \centering
     \begin{tabular}{cccccc} 
       $\textrm{B [G]}$&$\zeta$&$\textrm{G [T/m]}$&$\gamma^{1/3} n^2$&$\Delta E $&$\Delta \textrm{ [\%]}$\\
       \hline    \\[-10pt]
       $0.01$&$0.288$&$100$&$1.375$&$0$&\\
       $0.1$&$2.89$&$100$&$1.375$&$1.291$&$15.193$\\
       $1$&$28.87$&$100$&$1.375$&$14.421$&$1.476$\\
       $0.01$&$1.340$&$10$&$0.638$&$0.600$&$11.101$\\
       $0.1$&$13.40$&$10$&$0.638$&$6.694$&$0.103$\\
       $1$&$134.0$&$10$&$0.638$&$67.006$&$0.002$\\
       $10$&$1340$&$10$&$0.638$&$670.07$&$0.001$\\
       $0.01$&$6.220$&$1$&$0.296$&$3.107$&$0.104$\\
       $0.1$&$62.20$&$1$&$0.296$&$31.101$&$0.001$\\
       $1$&$622.0$&$1$&$0.296$&$311.022$&$0.000$\\
       \\[-10pt]
       \hline
     \end{tabular}
   \caption[Minimal Distance]{Minimal distance $\Delta E$ of the two uppermost surfaces of the $n=30$ manifold.
     $\Delta$ denotes the discrepancy between $\Delta E$ and the approximate predicted value for the distance, $\frac{\zeta}{2}$, 
     according to Eq.~(\ref{eq:mindist}).} 
   \label{t:mindist}
   \end{table}

To check the range of validity of our approximation, 
the minimal energetic distance between the two uppermost adiabatic surfaces 
in the $n=30$ manifold has been calculated for different parameters, 
subtracting the full 2D surfaces from each other, 
that have been obtained using the electronic Hamiltonian (\ref{eq:BOHe}). 
One finds the minimal distance to be located at the origin, as expected.
$\Delta$ in Tab.~\ref{t:mindist} denotes the relative deviation 
between the predicted (Eq.~(\ref{eq:mindist})) and the computed value in percent. 
It is small for large Ioffe field strengths $B$ and low field gradients $G$. 
Then we have~$\zeta \gg \gamma^{{1}/{3}}n^2$, the last term in the electronic Hamiltonian is negligible 
and our approximation that leads to~(\ref{eq:mindist}) is justifiable.

\subsection{High Gradients}                                                              \label{ch:highgradients}

A more complicated picture of the surface properties arises when the field gradients become larger. 
The last term in the electronic Hamiltonian, that accounts for finite size effects of the atom, 
\begin{equation}                                                 \label{eq:finitesize}
  \gamma^{\frac{1}{3}} (xyp_z+x S_x-y S_y) \; ,
\end{equation}
is no longer small compared to the others in equation (\ref{eq:BOHe}).
This results in modulations of the adiabatic surfaces we already spotted in the previous section, 
even though the term does not feature any dependency on $X$ and $Y$. 
These modulations lift the degeneracy that was found in the limit of small gradients. 
Their dependency on the c.m.~coordinates is introduced by the transformation $\mathcal U(X,Y)$ 
that diagonalizes the electronic problem (cf.~Sec.~\ref{s:aa}). 

\begin{figure}[tbp]
  \centering
  \includegraphics[width=7.5cm]{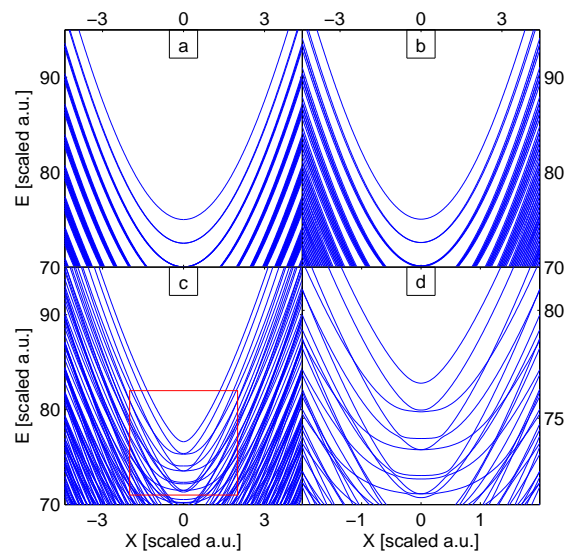} 
  \caption[Sections for increasing field gradient, $n=30$.]{Sections  ($Y=0$)
    through the adiabatic potential energy surfaces belonging to the 
    $n=30$ manifold of $^{87}$Rb for decreasing ratios $B/(G n^2)=\zeta/(\gamma^{{1}/{3}} n^2)$. 
    The influence of the Zeeman term in $\mathcal H_e$ (\ref{eq:BOHe}) is fixed ($\zeta=5$) 
    while $\gamma^{{1}/{3}} n^2$ increases. 
    (a) $B/(G n^2)=10 \leftrightarrow B=22.9$ mG, $G=4.81$ T/m; 
    (b) $B/(G n^2)=5  \leftrightarrow B=91.6$ mG, $G=38.5$ T/m; 
    (c) $B/(G n^2)=1  \leftrightarrow B=2.29$ G, $ G=4807$ T/m; 
    (d) draws the indicated region in (c) to a larger scale.}
  \label{f:zeta1_gammas}
\end{figure} 

In order to isolate the effect of the term (\ref{eq:finitesize}) on the adiabatic surfaces, 
we vary the scaling factor~$\gamma=GM$ by changing the field gradient~$G$, 
while keeping~$\zeta=BM^{{1}/{3}}G^{-{2}/{3}}$ constant. 
It is, for example, reasonable to demand $\zeta=5$ and 
to adjust the Ioffe field strength~$B$ to meet this condition. 
Fig.~\ref{f:zeta1_gammas} demonstrates the increasing influence of the interaction (\ref{eq:finitesize}) when $G$ is increased. 
The spectra are computed for the $n=30$ manifold of $^{87}$Rb, $\zeta=5$, 
while $G$ is varied from $4.8$ to $4800$~T/m. 
For small field gradients ((a), $B/(G n^2)=10$), 
the surfaces approach the shapes predicted in the limit addressed in the previous subsection (\ref{ch:regulatingcapacity}): 
The adiabatic surfaces with the same value of the magnetic moment $(m_l+2m_s)/2$ are approximately degenerate. 
The uppermost energy is the only non-degenerate one and 
to the corresponding eigenstate the quantum numbers $m_l=n-1$ and $m_s={1}/{2}$ can be assigned. 
An increasing field gradient lifts the degeneracy and groups of curves can be observed ((b), $B/(G n^2)=5$). 
The energetic distance between these groups stays tunable by the bias field strength, as we elucidated above (see Eq.~(\ref{eq:mindist})).
For even higher field gradients, the different parts of the electronic Hamiltonian are of comparable size 
and finite size effects substantially alter the shape of the energy surfaces ((c), (d), $B/(G n^2)=1$). 
Avoided level crossings appear and non-adiabatic transitions are likely to occur.
The uppermost energy surface, however, proves to be very robust when the field gradient is varied.
It is energetically well-isolated from the other adiabatic surfaces. 
Its distance to the surface, that is formed by the second highest eigenvalue, 
only decreases significantly when the ratio $B/(G n^2)$ approaches one ((c), (d)).
This holds true for the entire X-Y-plane. 
Inspecting the full uppermost surface one furthermore finds the azimuthal symmetry, 
that is found for large ratios $B/(G n^2)$ (see Sect.~\ref{ch:regulatingcapacity}), 
to be approximately conserved.

  \begin{figure}[bt!]
    \centering
    \includegraphics[width=8.5cm]{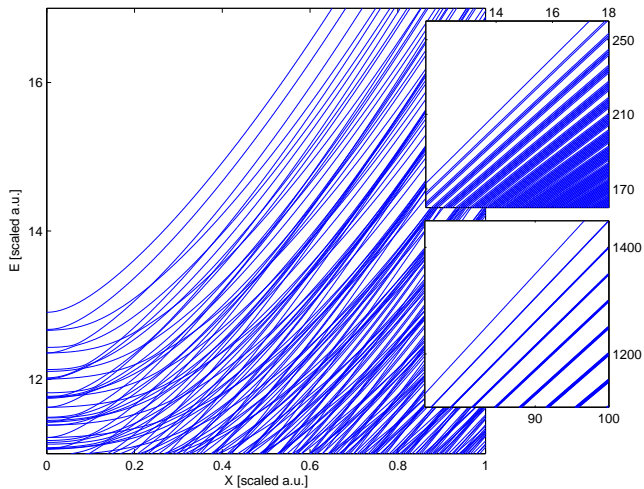}
    \caption[Section $0.01$ G $20$ T/m, $n=30$.]{
      Section through the $n=30$ manifold for a field strength of $0.01$ Gauss and a field gradient of $20$ T/m ($^{87}$Rb).
      A large number of avoided crossings can be observed.
      The uppermost curve, however, stays isolated from the other curves.
      The insets show the linear behavior of the surfaces far away from the $z$-axis.
    }
    \label{f:001G20T}
  \end{figure} 

Another example for the complicated structure of the adiabatic electronic energy surfaces is shown in Fig.~\ref{f:001G20T}.
The data are calculated for a Ioffe field strength of $0.01$~G and a field gradient of $20$~T/m. 
For these parameters, the contributions of all terms in the electronic Hamiltonian are of the same order of magnitude around $X=1$. 
One immediately notices the large number of avoided crossings between the surfaces. 
The uppermost curve however remains isolated from the rest of the curves. 
Far away from the trap center, i.e.~for large $\rho=\sqrt{X^2+Y^2}$, 
the coupling term in (\ref{eq:BOHe}), 
$X(\frac{1}{2}L_x+S_x) - Y(\frac{1}{2}L_y+S_y)$, becomes dominant. 
A Zeeman like splitting of the surfaces emerges, visible in the smaller graphs on the right.

\subsection{Electronic Wave Functions}

To characterize the electronic wave function $\varphi_\kappa(\bm r;\bm R)$,
that corresponds to the energy eigenvalues constituting the uppermost adiabatic surface, 
we analyze its radial extension, angular momentum and spin.
\begin{figure}[tbhp]
  \centering
  \includegraphics[width=8.5cm]{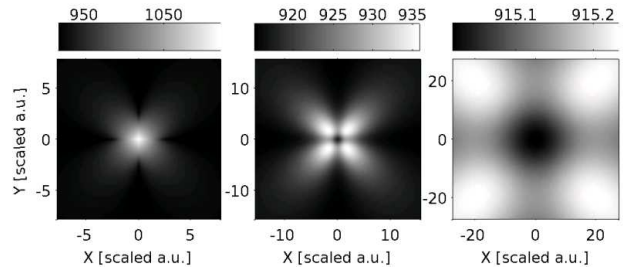} 
  \caption[Expectation value $\langle r \rangle_\varphi$]{
    Expectation value $\langle r\rangle_\varphi$ of the wave functions that correspond to the uppermost electronic energy surface 
    for $G=100$~T/m ($n=30$, $^{87}$Rb). 
    $B$ is varied yielding different values for the ratio $\zeta/\gamma^{{1}/{3}}n^2=B/Gn^2$: 
    (a) $0.01$~G $\rightarrow B/Gn^2=0.21 $~a.u., 
    (b) $0.1$~G $\rightarrow B/Gn^2=2.1 $, 
    (c) $1$~G $\rightarrow B/Gn^2=21 $.
    The depicted ranges of~$X$ and~$Y$ correspond to 30 characteristic lengths
    of the c.m.~motion in scaled units.} 
  \label{f:rexpall}
\end{figure}
The electronic wave function depends parametrically on the c.m.~position 
and is, in general, distorted compared to the field free case by the external magnetic field. 
This is reflected in the expectation value 
$\langle r \rangle_e (\bm R) = \langle \varphi_\kappa(\bm r;\bm R)| r |\varphi_\kappa(\bm r;\bm R)\rangle$ 
which is shown in Fig.~\ref{f:rexpall} for different ratios $B/(G n^2)$. 
The limits of the graphs with respect to $X$ and~$Y$ correspond to thirty characteristic lengths of the c.m.~motion. 
While keeping $G=100$~T/m, 
$B$ is increased for the different plots from left to right. 
For the smallest ratio under consideration ((a), $B/Gn^2<1$), 
a pronounced maximum of the expectation value $\langle r \rangle_e$ can be observed at the trap center. 
This maximum breaks up into four maxima arranged along the diagonals when the ratio is increased ((b), $B/Gn^2>1$), 
while the amplitude of the spatial variation of $\langle r \rangle_e$ decreases. 
For an even higher value of $B$ ((c), $B/Gn^2\gg 1$), only a marginal deviation from 
the hydrogenic field free value for the highest possible angular momentum quantum number remains 
(for $n=30$ one finds \hbox{$\langle r \rangle_H(n=30,l=29)=915$}). 
In the region of local homogeneity, 
where the magnetic field does not vary significantly over the extension of the electronic cloud 
(i.e.~far from the $z$-axis), 
the expectation value approaches the field free value in all subfigures that are shown in Fig.~\ref{f:rexpall}. 
In accordance with the abovementioned scaling property of the electronic Hamiltonian~$\mathcal H_e$, 
changing the field parameters while keeping the ratio $B/Gn^2$ unaltered only modifies the scale of the c.m.~coordinates, 
whereas the shape of the bright regions and the energy range of the eigenvalues are not changed. 

\begin{figure}[tbhp]
  \centering
  \includegraphics[width=8.5cm]{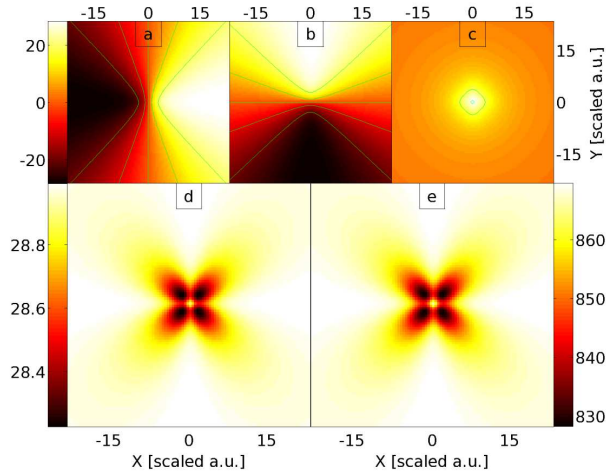} 
  \caption[]{
    Expectation values $\langle L_x\rangle$, $\langle L_y\rangle$ and $\langle L_z\rangle$ (a,b,c, respectively) 
    for a ratio $B/(G n^2)=2.1$ (Ioffe field $B=0.1$ G, gradient $G=100$ T/m, $^{87}$Rb, n=30). 
    In (d) the projection $\Pi$ of $\langle \bm {L_r}\rangle$ 
    onto the local magnetic field direction $\bm G$ is displayed. 
    It is close to the field free maximum value for the angular momentum projection, $m_{l,max}=n-1$. 
    Subplot (e) shows the spatial behavior of $\langle \bm L^2\rangle$.  
    The range of $X$ and $Y$ corresponds to 30 times the characteristic length of the c.m.\ motion. 
  }
  \label{f:Lrsmallratio}
\end{figure}

Let us study the angular momentum and its orientation. 
It is to be expected that for dominating Ioffe field, i.e.~for very large ratios $B/(G n^2)$, 
the expectation value of the angular momentum, 
$\langle\bm L_{\bm r}\rangle = (\langle L_{x}\rangle,\langle L_{y}\rangle,\langle L_{z}\rangle)$, 
is oriented in the Ioffe-field direction ($z$-axis). 
Since the Ioffe field in any case dominates around the origin, 
$\langle L_x \rangle$ and $\langle L_y \rangle$ are expected to vanish at $(X,Y)=(0,0)$ while  $\langle L_z \rangle$ becomes maximal. 
This behavior can be observed in Fig.~\ref{f:Lrsmallratio} where $\langle L_i\rangle$ are displayed (a,b,c) 
for $B=0.1$ G and $G=100$ T/m. 
These parameters yield $B/(G n^2)=2.1$. 
The alignment of $\langle\bm L_{\bm r}\rangle$ and the local field direction~$\bm{G}(X,Y)$ is found to be very good 
in the entire $X$-$Y$-plane (the maximum angle between the two is smaller than $3.6^\circ$). 
In subplot (d) we provide the spatial behavior of the projection of $\langle\bm L_{\bm r}\rangle$ onto this local field axis, 
$\Pi = \langle\bm L_{\bm r}\rangle\cdot {\bm{G}(\bm R)}/{|\bm{G}(\bm R)|}$.
In the local homogeneity limit, $\Pi$ approaches the maximal value for $\langle L_z\rangle$, 
namely $m_{l,max}=n-1$. 
In the same manner the expectation value $\langle \bm L^2\rangle$, which is displayed in subplot (e), 
converges to the maximal value, $l_{max}(l_{max}+1)=n(n-1)$.
Far from the $z$-axis, the uppermost surface hence corresponds to the circular state $|m_{l,max},l_{max}\rangle$.
The deviation of $\Pi$ and $\langle \bm L^2\rangle$ 
from the maximal values close to the $z$-axis reflect the admixture of states 
with lower quantum numbers $m$ and $l$ to the state of the uppermost surface. 

\begin{figure}[tbhp]
  \centering
  \includegraphics[width=5.5cm]{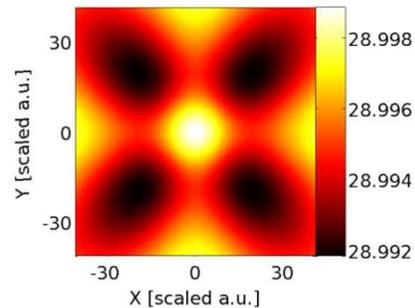}
  \caption[]{
    Spatial dependence of the projection $\Pi$ of  $\langle \bm {L_r}\rangle$ onto the local field axis for $B=1$ G 
    (all other parameters are the same as in Fig.~\ref{f:Lrsmallratio}). 
    For this ratio, $B/(G n^2)=21$, the deviations from the maximal value $m_{l,max}=n-1$ are marginal. 
    (Equally, $\langle \bm {L}^2\rangle\approx l_{max}(l_{max}+1)$, not shown.) 
  }
  \label{f:Lrbigratio}
\end{figure}

Increasing the applied Ioffe field by a factor of $10$ ($\rightarrow B/(G n^2)=21$), 
decreases the angle between $\langle\bm L_{\bm r}\rangle$ and $\bm{G}(X,Y)$ by a factor of $10^2$, 
i.e.\ a quasi perfect alignment is found. 
As can be seen in Fig.~\ref{f:Lrbigratio}, the projection $\Pi$ now only deviates marginally from $m_{l,max}$. 
Consequently, also $\langle \bm L^2\rangle$ exhibits only minor deviations from its maximum value in the whole $X$-$Y$-plane. 
For high ratios $B/(G n^2)$, the admixture is therefore marginal and 
one can in a very good approximation assume the electronic state in the uppermost surface 
to be the circular state $|m_{l,max},l_{max}\rangle$ for any c.m.\ position. 
Similar observations can be made considering the respective expectation values for the spin. 
For the parameters in Fig.~\ref{f:Lrbigratio}
the projection of $\langle\bm S\rangle$ onto $\bm G$ 
differs less than $10^{-4}$ from $1/2$.
The expectation values of the examined electronic observables converge to the field free values for increasing ratios $B/(G n^2)$. 

Our findings indicate that the electronic structure of the atom is barely changed in the limit of large ratios $B/(G n^2)$. 
The radiative lifetimes can hence be expected to differ only slightly from the field free ones~\cite{igor:pra72}.

\section{\label{s:cm}Quantized center of mass motion}

The energetically uppermost adiabatic electronic energy surface is the most appropriate to achieve confinement. 
It does not suffer a significant deformation when the field gradient is increased 
and it stays well isolated from lower surfaces for a wide range of parameters.
Large energetic distances to adjacent surfaces suppress nonadiabatic couplings 
(Eqs.~(\ref{eq:neglectdeltaT1}) and (\ref{eq:neglectdeltaT2})). 
In order to obtain the quantized c.m.~states 
we therefore solve the Schr\"odinger equation~(\ref{eq:BOcm}) for the c.m.~motion
in the uppermost surface $E_{2n^2}$ 
by discretizing the Hamiltonian on a grid. 
The wave function for the fully quantized state is hence composed of
the eigenfunction $|\varphi_\kappa(\bm r;\bm R)\rangle$ of the electronic Hamiltonian in equation (\ref{eq:internal}), 
the wave function for the center of mass motion in the $X$-$Y$ plane, $|\psi_\nu(\bm R)\rangle$,
and the plain wave in $Z$ direction, 
\begin{equation}
  |\Psi(\bm r,\bm R)\rangle =
  |\varphi_\kappa(\bm r;\bm R)\rangle
  \otimes |\psi_\nu(\bm R)\rangle
  \otimes | k_Z\rangle \; .
\end{equation}

\begin{figure}[tbhp]
  \centering
  \includegraphics[width=8.5cm]{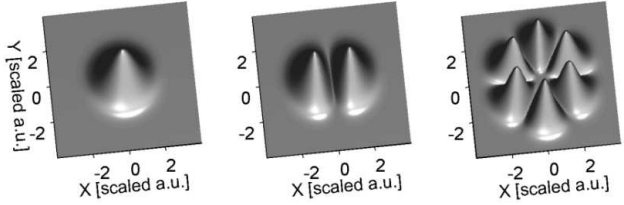}
  \caption[Probability densities of c.m.~states]{
    Probability densities of the ground state and the first and tenth excited states
    of the c.m.~motion in the uppermost adiabatic potential surface of the $n=30$ manifold of~$^{87}$Rb 
    (from left to right). 
    The Ioffe field strength is set to $B=0.1$ G and the field gradient is $G=10$ T/m.}
  \label{f:cm}
\end{figure}

In Fig.~\ref{f:cm} the probability densities of the ground state and two excited states of the c.m.~motion 
in the uppermost surface of the $n=30$ manifold of $^{87}$Rb are displayed. 
These densities reflect the spatial symmetries of the electronic Hamiltonian $\mathcal H_e$ (\ref{eq:BOHe})
and consequently those of the electronic energy surface. 
They are computed for a Ioffe field strength $B=0.1$~G and a field gradient of $G=10$~T/m, 
which yields $\zeta=13.4$ and $\zeta/ \gamma^{\frac{1}{3}}n^2= B/Gn^2= 21$. 
According to the discussion in Sec.~\ref{ch:regulatingcapacity}, 
the electronic surface then exhibits a harmonic behavior around the origin, 
and the system resembles the two dimensional isotropic harmonic oscillator 
in the potential  $E_h(X,Y) = (\zeta + \rho^2/2\zeta)\cdot {n}/{4}$ (cf.~Eq.~(\ref{eq:Eksmallrho}), $m_l=n-1$). 

The first two probability densities (from left to right) in Fig.~\ref{f:cm} 
explicitely demonstrate the analogy to the harmonic oscillator. 
The nodal structure of the tenth excited state is not due to a Cartesian product of 1D harmonic oscillators 
but a different combination of the harmonic oscillators in the corresponding degenerate subspace. 

The energies of the c.m.~wave functions in the approximative potential $E_h(X,Y)$ read 
\begin{equation}
  \epsilon_{h,\nu}= (N_1+N_2+1) \ \omega \; , \quad N_1,N_2=0,1,2\dots \; , 
\end{equation}
where $\omega^2 = {n}/{2\zeta}$, 
which are in very good agreement with the exact results in the regime where the electronic Hamiltonian (\ref{eq:finitesize}) is negligible. 
Within this approximation, 
the energy level spacing scales with the inverse square root of $\zeta$, 
$\Delta \epsilon_{h,\nu} = \omega \sim {1}/{\sqrt{\zeta}}$, 
whereas the energetic distance of adjacent surfaces scales linearly with~$\zeta$, see Eq.~(\ref{eq:mindist}).

\begin{figure}[bth]
  \centering
    \includegraphics[width=7.5cm]{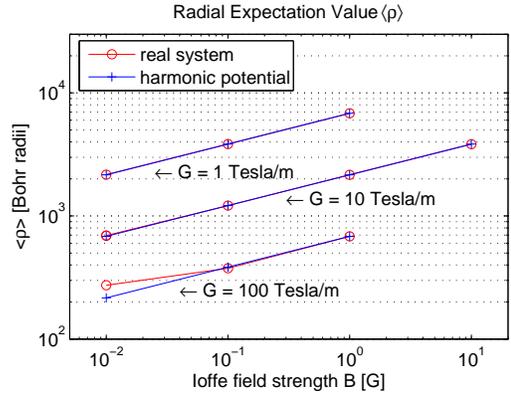}
    \caption[Expectation value $\langle \rho \rangle$.]{
      Double logarithmic plot of the expectation value $\langle \rho \rangle$ 
      for the c.m.~ground state (circles,~$\circ$) in the uppermost adiabatic energy surface ($n=30$, $^{87}$Rb).
      The corresponding expectation values for the c.m.~wave function in a perfectly harmonic potential are depicted for comparison~({\tiny{+}}).
    }     \label{f:rho}
\end{figure} 

To describe the properties of the compound quantized state, we analyze the extension of the center of mass motion, 
which can be measured by the expectation value
\begin{equation} 
  \langle\rho\rangle =
  \langle\psi_\nu(\bm R)| \  
  \sqrt{X^2+Y^2} \ 
  |\psi_\nu(\bm R)\rangle     \; ,
\end{equation}
and the mean distance of the core and the electron $\langle r\rangle$. 
Fig.~\ref{f:rho} presents the radial expectation value $\langle\rho\rangle$ in Bohr radii for the c.m.~ground state 
in the uppermost energy surface for different parameter sets of the magnetic field. 
For comparison, the expectation value of the c.m.~state in a perfectly harmonic potential, 
$\langle\rho\rangle_h = \frac{\sqrt{\pi}}{2} x_0 \sim \zeta^{{1}/{4}}$, is also depicted. 
The characteristic length of the c.m.~motion is $x_0 =  1/\sqrt{\omega}  = \sqrt[4]{{2\zeta}/{n}}$. 
(Due to the rescaling of the c.m.~coordinates with $\gamma^{-{1}/{3}}$ in Sec.~\ref{ch:ip} 
this is of the order of $1$ for a wide range of parameter sets \{$B$, $G$, $n$\} in scaled atomic units (cf.~Tab.~\ref{t:parameters}).) 
The expectation values for the real system, $\langle\rho\rangle$, 
deviate from the straight line formed by $\langle\rho\rangle_h$, 
as the ratio $B/G$ becomes very small. 
Hence, by choosing large gradients and appropriate bias fields, 
very tightly confining traps for highly excited atoms can be obtained 
($B=0.1$~G and $G=100$~T, for instance, give rise to a trap frequency of approximately $1.4$~MHz). 

\begin{figure}[tbhp]
  \centering

  \includegraphics[width=8.5cm]{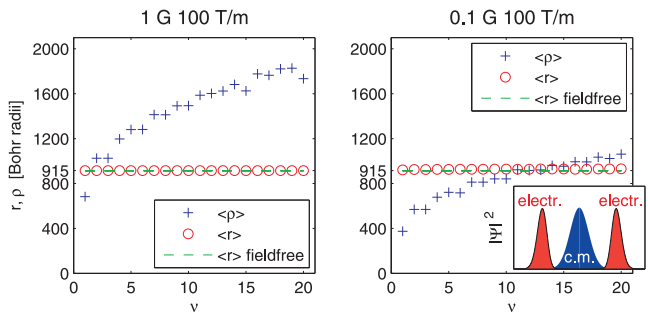} 
  \caption[Expectation values $\langle r \rangle$ and $\langle \rho \rangle$.]{
    Comparison of the mean extension of the c.m.~wave function, 
    $\langle\rho\rangle$, and the mean distance of the core and the electron, $\langle r \rangle$, for the $n=30$ manifold of $^{87}$Rb.
    }                                                                                                                       \label{f:r_rho}
\end{figure}

The mean distance of the Rydberg electron from the core $\langle r \rangle$ 
is calculated weighting that very quantity for a fixed c.m.~position $\langle r \rangle_e (X,Y)$, 
with the probability density of the c.m.~wave function: 
\begin{equation}
  \langle r \rangle 
  = \langle\psi_\nu(\bm R)| \  \langle\varphi_\kappa(\bm r;\bm R)| \ 
  r \
  |\varphi_\kappa(\bm r;\bm R)\rangle  \  |\psi_\nu(\bm R)\rangle     \; .
\end{equation}
It is depicted in Fig.~\ref{f:r_rho}, along with $\langle\rho\rangle$, versus the degree of excitation of the c.m.~motion~$\nu$. 
$\langle\rho\rangle$ and $\langle r \rangle$ are of comparable size 
due to the very tight confinement. 
For a Ioffe field strength of $B=0.1$ G and a field gradient of $G=100$ T/m, for instance, 
the ratio of $\langle \rho \rangle$ and~$\langle r \rangle$ for the ground state ($\nu=1$) is as small as 
${\langle \rho \rangle}/{\langle r \rangle} = 0.4$. 
The extension of the c.m.~wave function is thus smaller than the extension of the electronic cloud.  
This strongly supports the proposition that our Rydberg atoms cannot be considered as point-like particles. 

The expectation value $\langle r \rangle$ for the electron remains nearly constant as the degree of excitation increases, 
and it barely differs from the corresponding field free value (dashed line in Fig.~(\ref{f:r_rho})).
As indicated previously, we find the electron to be in the circular state with $m_l=n-1$, 
which features the smallest 
mean square deviation of the nucleus-electron separation 
$\langle r^2\rangle  - \langle r\rangle^2  = n^2(2n+1)/4$. 
It is therefore possible, that the c.m.~and the electronic wave function do not even overlap. 
This is indicated in the inset of the upper right plot in Fig.~\ref{f:r_rho} for $\nu=1$. 

\section{\label{s:c}Conclusion}

We have studied the quantum properties of ultracold Rydberg atoms in a Ioffe-Pritchard field configuration 
and find trapped c.m.~quantum states to be readily achievable. 
Our starting point is a two-body approach to the Rydberg atom. 
Relativistic effects and deviations of the core potential from the Coulomb-potential 
as well as diamagnetic interactions have not been taken into account, 
which is well justified a posteriori. 
Applying a spatially dependent unitary transformation 
and additionally exploiting the major mass difference of the electron and the core, 
we arrived at a two-particle Hamiltonian for highly excited atoms in an inhomogeneous field 
where the appearance of the coupling of the relative and c.m.~dynamics is simplified substantially. 
Thenceforward we have concentrated on the special case of a Ioffe-Pritchard trap. 
A symmetry analysis of the resulting Hamiltonian has been performed 
revealing seven discrete unitary and anti-unitary symmetries. 
Comparing the energetic contributions of the different interactions we find it legitimate 
to limit our considerations to a single $n$-manifold 
to solve the corresponding stationary Schr\"odinger equation. 
Consequently an adiabatic approach was applied. 
In the ultracold regime the Rydberg electron is much faster than the c.m.\ motion of the atom. 
This justifies an adiabatic separation of the internal (relative) and the external (c.m.) dynamics. 
The corresponding adiabatic electronic potential surfaces have been obtained by diagonalizing the electronic Hamiltonian matrix. 
In the limit of large ratios of Ioffe field strength and field gradient, $B/(G n^2)$, 
an approximate analytical expression for the adiabatic surfaces has been provided. 
In this limit, the surfaces arrange equidistantly and all but the uppermost surface are degenerate. 
The inter-surface distance is then proportional to the Ioffe field strength. 
The structure of the electronic surfaces becomes more complex when this ratio decreases. 
The shape of the uppermost surface and its energetic separation from others, however, 
prove very robust with respect to changes of the field parameters. 
We hence consider it the most appropriate to achieve confinement. 
Exploring the properties of the electronic wave functions 
we find that the expectation values approach the field free values 
when the ratio of the field strength and the field gradient,~$B/(G n^2)$, is increased. 
This indicates that, despite the strong localization of the c.m., 
the electronic structure of the atom is barely changed compared to the field free case. 
Examining the compound quantized states we have found a regime where the extension of the c.m.~wave function 
falls below the extension of the electronic cloud, 
i.e.~the c.m.~is stronger localized than the valence electron.
In this regime Rydberg atoms in inhomogeneous magnetic fields can therefore not be considered as point-like particles. 
We conclude that the Ioffe-Pritchard trap provides a strong confinement for Rydberg atoms in two dimensions 
that permits their trapping on a microscopic scale. 
For such a one-dimensional guide, 
a relatively weak longitudinal confinement along the $z$-axis could additionally be provided 
for a non-Helmholtz configuration by the quadratic term. 
As a natural enrichment of the system one could study many atoms in that guide. 
Challenging issues are to stabilize such a one-dimensional Rydberg gas or
to answer the question if it is feasible to use the strong Rydberg-Rydberg interaction to create a chain of trapped atoms \cite{mayle:1d} 
that could then serve as a tool for quantum information processing~\cite{lukin,sorensen,jaksch}
making use of the state dependent atom-atom interaction.

\section{Acknowledgment}

Financial support by the Deutsche Forschungsgemeinschaft is gratefully acknowledged. 

\bibliography{pra}

\begin{thebibliography}{33}
\expandafter\ifx\csname natexlab\endcsname\relax\def\natexlab#1{#1}\fi
\expandafter\ifx\csname bibnamefont\endcsname\relax
  \def\bibnamefont#1{#1}\fi
\expandafter\ifx\csname bibfnamefont\endcsname\relax
  \def\bibfnamefont#1{#1}\fi
\expandafter\ifx\csname citenamefont\endcsname\relax
  \def\citenamefont#1{#1}\fi
\expandafter\ifx\csname url\endcsname\relax
  \def\url#1{\texttt{#1}}\fi
\expandafter\ifx\csname urlprefix\endcsname\relax\def\urlprefix{URL }\fi
\providecommand{\bibinfo}[2]{#2}
\providecommand{\eprint}[2][]{\url{#2}}

\bibitem[{\citenamefont{Pethick}(2001)}]{pethick01}
\bibinfo{author}{\bibfnamefont{C.~J.} \bibnamefont{Pethick}},
  \emph{\bibinfo{title}{Bose-Einstein condensation in dilute gases}}
  (\bibinfo{publisher}{Cambridge University Press}, \bibinfo{year}{2001}).

\bibitem[{\citenamefont{Dalfovo et~al.}(1999)\citenamefont{Dalfovo, Giorgini,
  Pitaevskii, and Stringari}}]{dalfovo99}
\bibinfo{author}{\bibfnamefont{F.}~\bibnamefont{Dalfovo}},
  \bibinfo{author}{\bibfnamefont{S.}~\bibnamefont{Giorgini}},
  \bibinfo{author}{\bibfnamefont{L.~P.} \bibnamefont{Pitaevskii}},
  \bibnamefont{and}
  \bibinfo{author}{\bibfnamefont{S.}~\bibnamefont{Stringari}},
  \bibinfo{journal}{Rev.\ Mod.\ Phys.} \textbf{\bibinfo{volume}{71}},
  \bibinfo{pages}{463} (\bibinfo{year}{1999}).

\bibitem[{\citenamefont{Pitaevskii and Stringari}(2003)}]{pitaevskii03}
\bibinfo{author}{\bibfnamefont{L.}~\bibnamefont{Pitaevskii}} \bibnamefont{and}
  \bibinfo{author}{\bibfnamefont{S.}~\bibnamefont{Stringari}},
  \emph{\bibinfo{title}{Bose-Einstein Condensation}}
  (\bibinfo{publisher}{Oxford University Press}, \bibinfo{year}{2003}).

\bibitem[{\citenamefont{Gallagher}(1994)}]{gallagher}
\bibinfo{author}{\bibfnamefont{T.~F.} \bibnamefont{Gallagher}},
  \emph{\bibinfo{title}{Rydberg Atoms (Cambridge Monographs on Atomic,
  Molecular and Chemical Physics)}} (\bibinfo{publisher}{Cambridge University
  Press}, \bibinfo{year}{1994}).

\bibitem[{\citenamefont{Mourachko et~al.}(1998)\citenamefont{Mourachko,
  Comparat, de~Tomasi, Fioretti, Nosbaum, Akulin, and Pillet}}]{mourachko}
\bibinfo{author}{\bibfnamefont{I.}~\bibnamefont{Mourachko}},
  \bibinfo{author}{\bibfnamefont{D.}~\bibnamefont{Comparat}},
  \bibinfo{author}{\bibfnamefont{F.}~\bibnamefont{de~Tomasi}},
  \bibinfo{author}{\bibfnamefont{A.}~\bibnamefont{Fioretti}},
  \bibinfo{author}{\bibfnamefont{P.}~\bibnamefont{Nosbaum}},
  \bibinfo{author}{\bibfnamefont{V.~M.} \bibnamefont{Akulin}},
  \bibnamefont{and} \bibinfo{author}{\bibfnamefont{P.}~\bibnamefont{Pillet}},
  \bibinfo{journal}{Phys.\ Rev.\ Lett.} \textbf{\bibinfo{volume}{80}},
  \bibinfo{pages}{253} (\bibinfo{year}{1998}).

\bibitem[{\citenamefont{Pohl et~al.}(2003)\citenamefont{Pohl, Pattard, and
  Rost}}]{pohl1}
\bibinfo{author}{\bibfnamefont{T.}~\bibnamefont{Pohl}},
  \bibinfo{author}{\bibfnamefont{T.}~\bibnamefont{Pattard}}, \bibnamefont{and}
  \bibinfo{author}{\bibfnamefont{J.}~\bibnamefont{Rost}},
  \bibinfo{journal}{Phys.\ Rev.\ A} \textbf{\bibinfo{volume}{68}},
  \bibinfo{pages}{10703} (\bibinfo{year}{2003}).

\bibitem[{\citenamefont{Pohl et~al.}(2004)\citenamefont{Pohl, Pattard, and
  Rost}}]{pohl2}
\bibinfo{author}{\bibfnamefont{T.}~\bibnamefont{Pohl}},
  \bibinfo{author}{\bibfnamefont{T.}~\bibnamefont{Pattard}}, \bibnamefont{and}
  \bibinfo{author}{\bibfnamefont{J.}~\bibnamefont{Rost}},
  \bibinfo{journal}{Phys.\ Rev.\ Lett.} \textbf{\bibinfo{volume}{92}},
  \bibinfo{pages}{155003} (\bibinfo{year}{2004}).

\bibitem[{\citenamefont{Greene et~al.}(2000)\citenamefont{Greene, Dickinson,
  and Sadeghpour}}]{greene}
\bibinfo{author}{\bibfnamefont{C.}~\bibnamefont{Greene}},
  \bibinfo{author}{\bibfnamefont{A.}~\bibnamefont{Dickinson}},
  \bibnamefont{and}
  \bibinfo{author}{\bibfnamefont{H.}~\bibnamefont{Sadeghpour}},
  \bibinfo{journal}{Phys.\ Rev.\ Lett.} \textbf{\bibinfo{volume}{85}},
  \bibinfo{pages}{2458} (\bibinfo{year}{2000}).

\bibitem[{\citenamefont{Lesanovsky et~al.}(2006)\citenamefont{Lesanovsky,
  Schmelcher, and Sadeghpour}}]{igor:jpb39}
\bibinfo{author}{\bibfnamefont{I.}~\bibnamefont{Lesanovsky}},
  \bibinfo{author}{\bibfnamefont{P.}~\bibnamefont{Schmelcher}},
  \bibnamefont{and}
  \bibinfo{author}{\bibfnamefont{H.}~\bibnamefont{Sadeghpour}},
  \bibinfo{journal}{J. Phys. B: At. Mol. Phys.} \textbf{\bibinfo{volume}{39}},
  \bibinfo{pages}{L69} (\bibinfo{year}{2006}).

\bibitem[{\citenamefont{Singer et~al.}(2004)\citenamefont{Singer, Reetz-Lamour,
  Amthor, Marcassa, and Weidem\"uller}}]{singer}
\bibinfo{author}{\bibfnamefont{K.}~\bibnamefont{Singer}},
  \bibinfo{author}{\bibfnamefont{M.}~\bibnamefont{Reetz-Lamour}},
  \bibinfo{author}{\bibfnamefont{T.}~\bibnamefont{Amthor}},
  \bibinfo{author}{\bibfnamefont{L.~G.} \bibnamefont{Marcassa}},
  \bibnamefont{and}
  \bibinfo{author}{\bibfnamefont{M.}~\bibnamefont{Weidem\"uller}},
  \bibinfo{journal}{Phys.\ Rev.\ Lett.} \textbf{\bibinfo{volume}{93}},
  \bibinfo{pages}{163001} (\bibinfo{year}{2004}).

\bibitem[{\citenamefont{{Tong} et~al.}(2004)\citenamefont{{Tong}, {Farooqi},
  {Stanojevic}, {Krishnan}, {Zhang}, {C{\^o}t{\'e}}, {Eyler}, and
  {Gould}}}]{tong}
\bibinfo{author}{\bibfnamefont{D.}~\bibnamefont{{Tong}}},
  \bibinfo{author}{\bibfnamefont{S.~M.} \bibnamefont{{Farooqi}}},
  \bibinfo{author}{\bibfnamefont{J.}~\bibnamefont{{Stanojevic}}},
  \bibinfo{author}{\bibfnamefont{S.}~\bibnamefont{{Krishnan}}},
  \bibinfo{author}{\bibfnamefont{Y.~P.} \bibnamefont{{Zhang}}},
  \bibinfo{author}{\bibfnamefont{R.}~\bibnamefont{{C{\^o}t{\'e}}}},
  \bibinfo{author}{\bibfnamefont{E.~E.} \bibnamefont{{Eyler}}},
  \bibnamefont{and} \bibinfo{author}{\bibfnamefont{P.~L.}
  \bibnamefont{{Gould}}}, \bibinfo{journal}{Phys.\ Rev.\ Lett.}
  \textbf{\bibinfo{volume}{93}}, \bibinfo{pages}{063001}
  (\bibinfo{year}{2004}).

\bibitem[{\citenamefont{Ryabtsev et~al.}(2005)\citenamefont{Ryabtsev,
  Tretyakov, and Beterov}}]{ryabtsev}
\bibinfo{author}{\bibfnamefont{I.~I.} \bibnamefont{Ryabtsev}},
  \bibinfo{author}{\bibfnamefont{D.~B.} \bibnamefont{Tretyakov}},
  \bibnamefont{and} \bibinfo{author}{\bibfnamefont{I.~I.}
  \bibnamefont{Beterov}}, \bibinfo{journal}{J.~Phys.~B:~At.~Mol.~Opt.~Phys.}
  \textbf{\bibinfo{volume}{38}}, \bibinfo{pages}{S421} (\bibinfo{year}{2005}).

\bibitem[{\citenamefont{Lukin et~al.}(2001)\citenamefont{Lukin, Fleischhauer,
  C\^{o}t\'{e}, Duan, Jaksch, Cirac, and Zoller}}]{lukin}
\bibinfo{author}{\bibfnamefont{M.~D.} \bibnamefont{Lukin}},
  \bibinfo{author}{\bibfnamefont{M.}~\bibnamefont{Fleischhauer}},
  \bibinfo{author}{\bibfnamefont{R.}~\bibnamefont{C\^{o}t\'{e}}},
  \bibinfo{author}{\bibfnamefont{L.~M.} \bibnamefont{Duan}},
  \bibinfo{author}{\bibfnamefont{D.}~\bibnamefont{Jaksch}},
  \bibinfo{author}{\bibfnamefont{J.~I.} \bibnamefont{Cirac}}, \bibnamefont{and}
  \bibinfo{author}{\bibfnamefont{P.}~\bibnamefont{Zoller}},
  \bibinfo{journal}{Phys.\ Rev.\ Lett.} \textbf{\bibinfo{volume}{87}},
  \bibinfo{pages}{37901} (\bibinfo{year}{2001}).

\bibitem[{\citenamefont{Choi et~al.}(2005)\citenamefont{Choi, Guest, Povilus,
  Hansis, and Raithel}}]{choi}
\bibinfo{author}{\bibfnamefont{J.-H.} \bibnamefont{Choi}},
  \bibinfo{author}{\bibfnamefont{J.~R.} \bibnamefont{Guest}},
  \bibinfo{author}{\bibfnamefont{A.~P.} \bibnamefont{Povilus}},
  \bibinfo{author}{\bibfnamefont{E.}~\bibnamefont{Hansis}}, \bibnamefont{and}
  \bibinfo{author}{\bibfnamefont{G.}~\bibnamefont{Raithel}},
  \bibinfo{journal}{Phys.\ Rev.\ Lett.} \textbf{\bibinfo{volume}{95}},
  \bibinfo{pages}{243001} (\bibinfo{year}{2005}).

\bibitem[{\citenamefont{Choi et~al.}(2006)\citenamefont{Choi, Guest, and
  Raithel}}]{choi:epjd}
\bibinfo{author}{\bibfnamefont{J.-H.} \bibnamefont{Choi}},
  \bibinfo{author}{\bibfnamefont{J.~R.} \bibnamefont{Guest}}, \bibnamefont{and}
  \bibinfo{author}{\bibfnamefont{G.}~\bibnamefont{Raithel}},
  \bibinfo{journal}{Eur.\ Phys.\ J.\ D} \textbf{\bibinfo{volume}{40}},
  \bibinfo{pages}{19} (\bibinfo{year}{2006}).

\bibitem[{\citenamefont{Lesanovsky
  et~al.}(2004{\natexlab{a}})\citenamefont{Lesanovsky, Schmiedmayer, and
  Schmelcher}}]{igor:epl}
\bibinfo{author}{\bibfnamefont{I.}~\bibnamefont{Lesanovsky}},
  \bibinfo{author}{\bibfnamefont{J.}~\bibnamefont{Schmiedmayer}},
  \bibnamefont{and}
  \bibinfo{author}{\bibfnamefont{P.}~\bibnamefont{Schmelcher}},
  \bibinfo{journal}{Europhys.\ Lett.} \textbf{\bibinfo{volume}{65}},
  \bibinfo{pages}{478} (\bibinfo{year}{2004}{\natexlab{a}}).

\bibitem[{\citenamefont{Lesanovsky et~al.}(2005)\citenamefont{Lesanovsky,
  Schmiedmayer, and Schmelcher}}]{igor:jpb}
\bibinfo{author}{\bibfnamefont{I.}~\bibnamefont{Lesanovsky}},
  \bibinfo{author}{\bibfnamefont{J.}~\bibnamefont{Schmiedmayer}},
  \bibnamefont{and}
  \bibinfo{author}{\bibfnamefont{P.}~\bibnamefont{Schmelcher}},
  \bibinfo{journal}{J.~Phys.\ B} \textbf{\bibinfo{volume}{38}},
  \bibinfo{pages}{151} (\bibinfo{year}{2005}).

\bibitem[{\citenamefont{Lesanovsky and
  Schmelcher}(2005{\natexlab{a}})}]{igor:prl}
\bibinfo{author}{\bibfnamefont{I.}~\bibnamefont{Lesanovsky}} \bibnamefont{and}
  \bibinfo{author}{\bibfnamefont{P.}~\bibnamefont{Schmelcher}},
  \bibinfo{journal}{Phys.\ Rev.\ Lett.} \textbf{\bibinfo{volume}{95}},
  \bibinfo{pages}{053001} (\bibinfo{year}{2005}{\natexlab{a}}).

\bibitem[{\citenamefont{Lesanovsky and
  Schmelcher}(2005{\natexlab{b}})}]{igor:pra72}
\bibinfo{author}{\bibfnamefont{I.}~\bibnamefont{Lesanovsky}} \bibnamefont{and}
  \bibinfo{author}{\bibfnamefont{P.}~\bibnamefont{Schmelcher}},
  \bibinfo{journal}{Phys.\ Rev.\ A} \textbf{\bibinfo{volume}{72}},
  \bibinfo{pages}{53410} (\bibinfo{year}{2005}{\natexlab{b}}).

\bibitem[{\citenamefont{Hezel et~al.}(2006)\citenamefont{Hezel, Lesanovsky, and
  Schmelcher}}]{hezel:prl}
\bibinfo{author}{\bibfnamefont{B.}~\bibnamefont{Hezel}},
  \bibinfo{author}{\bibfnamefont{I.}~\bibnamefont{Lesanovsky}},
  \bibnamefont{and}
  \bibinfo{author}{\bibfnamefont{P.}~\bibnamefont{Schmelcher}},
  \bibinfo{journal}{Phys.\ Rev.\ Lett.} \textbf{\bibinfo{volume}{97}},
  \bibinfo{pages}{223001} (\bibinfo{year}{2006}).

\bibitem[{\citenamefont{Bethe and Salpeter}(1977)}]{bethe}
\bibinfo{author}{\bibfnamefont{H.~A.} \bibnamefont{Bethe}} \bibnamefont{and}
  \bibinfo{author}{\bibfnamefont{E.~E.} \bibnamefont{Salpeter}},
  \emph{\bibinfo{title}{Quantum Mechanics of One- and Two-Electron Atoms}}
  (\bibinfo{publisher}{Springer}, \bibinfo{year}{1977}).

\bibitem[{\citenamefont{Lesanovsky}(2006)}]{igor:diss}
\bibinfo{author}{\bibfnamefont{I.}~\bibnamefont{Lesanovsky}}, Ph.D. thesis
  (\bibinfo{year}{2006}).

\bibitem[{\citenamefont{Dippel et~al.}(1994)\citenamefont{Dippel, Schmelcher,
  and Cederbaum}}]{dippel}
\bibinfo{author}{\bibfnamefont{O.}~\bibnamefont{Dippel}},
  \bibinfo{author}{\bibfnamefont{P.}~\bibnamefont{Schmelcher}},
  \bibnamefont{and} \bibinfo{author}{\bibfnamefont{L.~S.}
  \bibnamefont{Cederbaum}}, \bibinfo{journal}{Phys.\ Rev.\ A}
  \textbf{\bibinfo{volume}{49}}, \bibinfo{pages}{4415} (\bibinfo{year}{1994}).

\bibitem[{\citenamefont{Schmelcher and Cederbaum}(1992)}]{peter:pla}
\bibinfo{author}{\bibfnamefont{P.}~\bibnamefont{Schmelcher}} \bibnamefont{and}
  \bibinfo{author}{\bibfnamefont{L.~S.} \bibnamefont{Cederbaum}},
  \bibinfo{journal}{Phys. Lett. A} \textbf{\bibinfo{volume}{164}},
  \bibinfo{pages}{305} (\bibinfo{year}{1992}).

\bibitem[{\citenamefont{Avron et~al.}(1978)\citenamefont{Avron, Herbst, and
  Simon}}]{avron}
\bibinfo{author}{\bibfnamefont{J.~E.} \bibnamefont{Avron}},
  \bibinfo{author}{\bibfnamefont{I.~W.} \bibnamefont{Herbst}},
  \bibnamefont{and} \bibinfo{author}{\bibfnamefont{B.}~\bibnamefont{Simon}},
  \bibinfo{journal}{Annals of Physics} \textbf{\bibinfo{volume}{114}},
  \bibinfo{pages}{431} (\bibinfo{year}{1978}).

\bibitem[{\citenamefont{Schmelcher and Cederbaum}(1993)}]{peter:cpl208}
\bibinfo{author}{\bibfnamefont{P.}~\bibnamefont{Schmelcher}} \bibnamefont{and}
  \bibinfo{author}{\bibfnamefont{L.~S.} \bibnamefont{Cederbaum}},
  \bibinfo{journal}{Chem. Phys. Lett.} \textbf{\bibinfo{volume}{208}},
  \bibinfo{pages}{548} (\bibinfo{year}{1993}).

\bibitem[{\citenamefont{Schmelcher et~al.}(1994)\citenamefont{Schmelcher,
  Cederbaum, and Kappes}}]{peter:ctqc}
\bibinfo{author}{\bibfnamefont{P.}~\bibnamefont{Schmelcher}},
  \bibinfo{author}{\bibfnamefont{L.~S.} \bibnamefont{Cederbaum}},
  \bibnamefont{and} \bibinfo{author}{\bibfnamefont{U.}~\bibnamefont{Kappes}},
  \emph{\bibinfo{title}{Molecules in Magnetic Fields: Fundamental Aspects}}
  (\bibinfo{publisher}{Kluwer Academic Publishers}, \bibinfo{year}{1994}), pp.
  \bibinfo{pages}{1--51}, Conceptual Trends in Quantum Chemistry.

\bibitem[{\citenamefont{Mewes et~al.}(1996)\citenamefont{Mewes, Andrews, van
  Druten, Kurn, Durfee, and Ketterle}}]{cloverleaf}
\bibinfo{author}{\bibfnamefont{M.-O.} \bibnamefont{Mewes}},
  \bibinfo{author}{\bibfnamefont{M.~R.} \bibnamefont{Andrews}},
  \bibinfo{author}{\bibfnamefont{N.~J.} \bibnamefont{van Druten}},
  \bibinfo{author}{\bibfnamefont{D.~M.} \bibnamefont{Kurn}},
  \bibinfo{author}{\bibfnamefont{D.~S.} \bibnamefont{Durfee}},
  \bibnamefont{and} \bibinfo{author}{\bibfnamefont{W.}~\bibnamefont{Ketterle}},
  \bibinfo{journal}{Phys.\ Rev.\ Lett.} \textbf{\bibinfo{volume}{77}},
  \bibinfo{pages}{416} (\bibinfo{year}{1996}).

\bibitem[{\citenamefont{Folman et~al.}(2002)\citenamefont{Folman, Kr\"uger,
  Schmiedmayer, Denschlag, and Henkel}}]{folman}
\bibinfo{author}{\bibfnamefont{R.}~\bibnamefont{Folman}},
  \bibinfo{author}{\bibfnamefont{P.}~\bibnamefont{Kr\"uger}},
  \bibinfo{author}{\bibfnamefont{J.}~\bibnamefont{Schmiedmayer}},
  \bibinfo{author}{\bibfnamefont{J.}~\bibnamefont{Denschlag}},
  \bibnamefont{and} \bibinfo{author}{\bibfnamefont{C.}~\bibnamefont{Henkel}},
  \bibinfo{journal}{Adv.~At.~Mol.~Opt.~Phys.} \textbf{\bibinfo{volume}{48}},
  \bibinfo{pages}{263} (\bibinfo{year}{2002}).

\bibitem[{\citenamefont{Lesanovsky
  et~al.}(2004{\natexlab{b}})\citenamefont{Lesanovsky, Schmiedmayer, and
  Schmelcher}}]{igor:pra70:4}
\bibinfo{author}{\bibfnamefont{I.}~\bibnamefont{Lesanovsky}},
  \bibinfo{author}{\bibfnamefont{J.}~\bibnamefont{Schmiedmayer}},
  \bibnamefont{and}
  \bibinfo{author}{\bibfnamefont{P.}~\bibnamefont{Schmelcher}},
  \bibinfo{journal}{Phys.\ Rev.\ A} \textbf{\bibinfo{volume}{70}},
  \bibinfo{pages}{43409} (\bibinfo{year}{2004}{\natexlab{b}}).

\bibitem[{\citenamefont{Mayle et~al.}(2007)\citenamefont{Mayle, Hezel,
  Lesanovsky, and Schmelcher}}]{mayle:1d}
\bibinfo{author}{\bibfnamefont{M.}~\bibnamefont{Mayle}},
  \bibinfo{author}{\bibfnamefont{B.}~\bibnamefont{Hezel}},
  \bibinfo{author}{\bibfnamefont{I.}~\bibnamefont{Lesanovsky}},
  \bibnamefont{and}
  \bibinfo{author}{\bibfnamefont{P.}~\bibnamefont{Schmelcher}}
  (\bibinfo{year}{2007}), \eprint{arXiv:0704.2299v1 [physics.atom-ph]}.

\bibitem[{\citenamefont{Sorensen et~al.}(2004)\citenamefont{Sorensen, van~der
  Wal, Childress, and Lukin}}]{sorensen}
\bibinfo{author}{\bibfnamefont{A.~S.} \bibnamefont{Sorensen}},
  \bibinfo{author}{\bibfnamefont{C.~H.} \bibnamefont{van~der Wal}},
  \bibinfo{author}{\bibfnamefont{L.~I.} \bibnamefont{Childress}},
  \bibnamefont{and} \bibinfo{author}{\bibfnamefont{M.~D.} \bibnamefont{Lukin}},
  \bibinfo{journal}{Phys.\ Rev.\ Lett.} \textbf{\bibinfo{volume}{92}},
  \bibinfo{pages}{63601} (\bibinfo{year}{2004}).

\bibitem[{\citenamefont{Jaksch et~al.}(2000)\citenamefont{Jaksch, Cirac,
  Zoller, Rolston, C\^{o}t\'{e}, and Lukin}}]{jaksch}
\bibinfo{author}{\bibfnamefont{D.}~\bibnamefont{Jaksch}},
  \bibinfo{author}{\bibfnamefont{J.~I.} \bibnamefont{Cirac}},
  \bibinfo{author}{\bibfnamefont{P.}~\bibnamefont{Zoller}},
  \bibinfo{author}{\bibfnamefont{S.~L.} \bibnamefont{Rolston}},
  \bibinfo{author}{\bibfnamefont{R.}~\bibnamefont{C\^{o}t\'{e}}},
  \bibnamefont{and} \bibinfo{author}{\bibfnamefont{M.~D.} \bibnamefont{Lukin}},
  \bibinfo{journal}{Phys.\ Rev.\ Lett.} \textbf{\bibinfo{volume}{85}},
  \bibinfo{pages}{2208} (\bibinfo{year}{2000}).

\end{thebibliography}

\end{document}